\pdfoutput=1
\documentclass[12pt,a4paper]{article}
\usepackage[dvipdfmx,hiresbb]{graphicx}
\usepackage{amsmath,amssymb,enumerate}
\usepackage{subfigure, url, ulem}
\usepackage{cite,booktabs}
\usepackage{comment}
\usepackage{mediabb}

\usepackage{color}

\def\a{\alpha}
\def\b{\beta}

\def\d{\delta}

\def\g{\gamma}

\def\k{\kappa}
\def\l{\lambda}
\def\m{\mu}
\def\n{\nu}

\def\s{\sigma}
\def\t{\tau}

\def\D{\Delta}
\def\G{\Gamma}

\def\beq{\begin{eqnarray}}
\def\eeq{\end{eqnarray}}

\newcommand{\lsim}{ \mathop{}_{\textstyle \sim}^{\textstyle <} }

\newcommand{\GEV}{ {\rm GeV} }

\begin{document}
\baselineskip 0.7cm
\begin{titlepage}

\begin{flushright}
KEK-TH-1803
\end{flushright}

\vskip 1.35cm
\begin{center}
{\large \bf 
Lepton-Specific Two Higgs Doublet Model \\
as a Solution of Muon $g-2$ Anomaly
}

\vskip 1.2cm
Tomohiro Abe$^1$, Ryosuke Sato$^1$ and Kei Yagyu$^2$
\vskip 0.4cm

{\it
$^1$Institute of Particle and Nuclear Studies,\\
High Energy Accelerator Research Organization (KEK),\\
Tsukuba 305-0801, Japan\\
$^2$School of Physics and Astronomy, University of Southampton,\\
Southampton, SO17 1BJ, United Kingdom
}

\vskip 1.5cm

\abstract{
We discuss the Type-X (lepton-specific) two Higgs doublet model 
as a solution of the anomaly of the muon $g-2$.
We consider various experimental constraints on the parameter space 
such as direct searches for extra Higgs bosons at the LEP II and the LHC Run-I,  
electroweak precision observables, the decay of $B_s \to \m^+\m^-$, and the leptonic decay of the tau lepton.
We find that the measurement of the tau decay provides the most important constraint, 
which excludes the parameter region that can explain the muon $g-2$ anomaly at the 1$\sigma$ level.  
We then discuss the phenomenology of extra Higgs bosons and the standard model-like Higgs boson ($h$) 
to probe the scenario favored by the $g-2$ data at the collider experiments.
We find that the $4\t$, $3\t$ and $4\t + W/Z$ signatures are expected as the main signal of the extra Higgs bosons at the LHC. 
In addition, we clarify that the value of the $h\t\t$ coupling is predicted to be the standard model value times about $-1.6$ to $-1.0$, and 
the branching fraction of the $h\to \gamma\gamma$ mode deviates from the standard model prediction by $-30\%$ to $-15\%$.  
Furthermore, we find that the exotic decay mode, $h$ decaying into the $Z$ boson and a light CP-odd scalar boson, 
is allowed, and its branching fraction  can be a few percent. 
These deviations in the property of $h$ will be tested by the precision measurements at future collider experiments.
}
\end{center}
\end{titlepage}
\setcounter{page}{2}

\section{Introduction}

The anomalous magnetic moment of the muon $a_\m \equiv (g-2)/2$, so-called muon $g-2$, is a very precisely measured observable.
The latest measurement of $a_\m$ by the E821 collaboration \cite{Bennett:2006fi} gives
\begin{align}
a_\m^{\rm exp} = 11~659~208.0~(5.4)(3.3) \times 10^{-10}.
\end{align}
As it has been well known that 
there is a discrepancy between the experimental value and the prediction of the standard model (SM).
According to the calculation evaluated in Refs.~\cite{Davier:2010nc, Hagiwara:2011af}
\begin{align}
a_\m^{\rm exp} - a_\m^{\rm SM} &= (28.7 \pm 8.0) \times 10^{-10}, \qquad (\textrm{Davier~et.~al.}) \nonumber\\
a_\m^{\rm exp} - a_\m^{\rm SM} &= (26.1 \pm 8.0) \times 10^{-10}, \qquad (\textrm{Hagiwara~et.~al.}) \nonumber
\end{align}
the discrepancy is more than the 3$\sigma$ level, which can be considered as an indirect evidence of the existence of a new physics model. 
This discrepancy will be further probed at Fermilab \cite{Grange:2015fou} and J-PARC \cite{Iinuma:2011zz} in the near future.
Since the size of the deviation is the same order as
the electroweak contribution $a_\m^{\rm EW} = 15.4 \times 10^{-10}$ \cite{Czarnecki:2002nt},
we expect that new physics exists at the electroweak scale if the strength of new interactions is as large as that of the weak interaction. 
In such a new physics scenario, 
new particles are expected to be light enough to be directly discovered at the LHC. 
Therefore, it is quite interesting to consider models beyond the SM as a solution of the muon $g-2$ anomaly.

Among various models which can explain the anomaly (for a review, {\it e.g.}, see Ref.~\cite{Jegerlehner:2009ry}), 
two Higgs doublet models (2HDMs) give simple solutions.
In 2HDMs, there are extra Higgs bosons ($H$, $A$, and $H^{\pm}$)
in addition to the SM-like Higgs boson ($h$), and they can give new contributions to $a_\m$.
Usually, a softly-broken discrete $Z_2$ symmetry is imposed~\cite{GW} to avoid flavor changing neutral current (FCNC) processes at the tree level. 
Under the $Z_2$ symmetry, four independent types of Yukawa interactions are allowed depending on the assignment of the $Z_2$ charge to the SM fermion~\cite{Barger,Grossman}, 
which are called as Type-I, Type-II, Type-X (or lepton specific) and
Type-Y (or flipped)~\cite{Aoki:2009ha}. 
In all the types of Yukawa interactions, the lepton couplings to the
extra Higgs bosons can be sizable enough to explain $a_{\mu}$. In the
Type-I and Type-Y 2HDMs, however, the top Yukawa coupling also becomes large
together with the enhancement of the lepton couplings. This is disfavored
from the view point of perturbativity. Thus, the Type-II and Type-X 2HDMs are suitable 
to solve the muon $g-2$ anomaly.

The muon $g-2$ has been calculated in a number of papers within 2HDMs
\cite{Dedes:2001nx,Chang:2000ii,Kingman1,Krawczyk:2001pe,Kingman2,Wu:2001vq,Gunion:2008dg,Cao:2009as,Chun,Wang:2014sda, Ilisie:2015tra}. 
In the early 2000s, this was calculated at the one-loop level in the Type-II 2HDM in Ref.~\cite{Dedes:2001nx}. 
After that, it was pointed out in Refs.~\cite{Chang:2000ii,Kingman1} that the two-loop Barr-Zee type diagrams~\cite{Barr-Zee1,Barr-Zee2} 
give a significant contribution to $a_\mu$
if a mass of $A$ is $\mathcal{O}(10\text{-}100)$ GeV and if there is large $Ab\bar{b}$ or $A\tau^+\tau^-$ couplings. 
In Ref.~\cite{Cao:2009as}, the implication on collider signatures
was discussed in the Type-X 2HDM, namely, the $h\to AA\to 4\tau$ process can be important in the favored parameter region by $a_\mu$. 
After the discovery of the Higgs boson at the LHC~\cite{Aad:2012tfa, Chatrchyan:2012ufa},
the muon $g-2$ has been reanalyzed 
by taking into account the Higgs boson search data in addition to the previous experimental constraints \cite{Chun,Wang:2014sda, Ilisie:2015tra}.  
Furthermore, the recent observation of ${\rm Br}(B_s \to \m^+ \m^-)$ at the LHC \cite{CMSandLHCbCollaborations:2013pla} gives
a new constraint on the parameter space of 2HDMs~\cite{Wang:2014sda}.

The difference between the Type-II and Type-X 2HDMs is the quark
couplings to the extra Higgs bosons.
In the Type-II 2HDM, both the lepton and down-type quark couplings are enhanced simultaneously,
and thus the model is severely constrained by flavor physics and direct searches of the extra Higgs bosons.
On the other hand, in the Type-X 2HDM, the quark couplings to the extra
Higgs bosons are suppressed when the lepton couplings are enhanced. 
Thus, the constraints are weaker than those in the Type-II 2HDM. 
In fact, it was clarified in Refs.~\cite{Chun,Wang:2014sda} that 
only the Type-X 2HDM can solve the muon $g-2$ anomaly with satisfying the current experimental data\footnote{
In addition to the muon $g-2$ anomaly,
there are several other motivations for this model.
For example, see Refs.~\cite{Ma:2002pf,Aoki:2008av}.
}.

Another important constraint comes from the lepton flavor physics.
In the Type-X 2HDM, the constraint from the leptonic $\tau$
decay~\cite{Hollik,Krawczyk:2004na, Aoki:2009ha, Logan:2009uf} 
gives a severe constraint on the parameter space favored to explain the $g-2$
anomaly because of the lepton coupling enhancements.
However, this important constraint has not been included in the previous analyses. 
Therefore, in this paper, we calculate the leptonic $\tau$ decay and the $Z\tau\tau$ vertex at the one-loop level in the Type-X 2HDM 
in order to compare the precise experimental measurements. 
We then investigate the favored parameter region by $a_\mu$
under these constraints in addition to those already known. 
Furthermore, we evaluate the running of the scalar quartic couplings by renormalization group equations (RGEs), and 
require that the couplings do not become too large up to a certain energy scale, for example 10 TeV. 
We find that extra Higgs boson loop contributions can reduce the discrepancy in $a_\mu$ to be 2$\sigma$ level, but not less than 1$\sigma$ level. 
We then study the collider phenomenology in the favored parameter region.

This paper is organized as follows. 
In Sec.~\ref{sec:2hdm_x}, we define the Lagrangian of the 2HDM, and derive 
the Higgs boson couplings with the gauge bosons and the fermions. 
In Sec.~\ref{sec:constraints}, we discuss constraints 
from direct searches for the extra Higgs bosons at LEP II and the LHC Run-I,  
electroweak precision observables, the decay of $B_s \to \m^+\m^-$, the leptonic $\tau$ decay, and the triviality bound. 
In Sec.~\ref{sec:muon_g-2}, we show the favored parameter regions by the muon $g-2$ anomaly. 
In Sec.~\ref{sec:phenomenology}, we discuss the collider phenomenology of the extra Higgs
bosons at the LHC, the deviations in the SM-like Higgs boson $h$ couplings, and the decay branching fractions of $h$. 
We also discuss the exotic decay mode: $h\to ZA$.
Conclusion is given in Sec.~\ref{sec:conclusion}. 
In Appendix,  we present the expressions for the decay rates of extra Higgs bosons and  
those for the parton level cross sections for the production of extra Higgs bosons at the LHC.

\section{The 2HDMs}\label{sec:2hdm_x}
\begin{table}
\centering
\begin{tabular}{|c||cc|ccc|c||c|c|c|}
\hline
& $H_1$ & $H_2$ & $u_R^c$ & $d_R^c$ & $\ell_R^c$ & $Q_L,L_L$ & $\xi_u$ & $\xi_d$ & $\xi_\ell$\\\hline\hline
Type-I  & $+$ & $-$	& $-$ & $-$ & $-$ & $+$    & $\cot\b$ & $\cot\b$ & $\cot\b$ \\\hline
Type-II & $+$ & $-$	& $-$ & $+$ & $+$ & $+$    & $\cot\b$ & $-\tan\b$ & $-\tan\b$ \\\hline
Type-X  & $+$ & $-$	& $-$ & $-$ & $+$ & $+$    & $\cot\b$ & $\cot\b$ & $-\tan\b$ \\\hline
Type-Y  & $+$ & $-$	& $-$ & $+$ & $-$ & $+$    & $\cot\b$ & $-\tan\b$ & $\cot\b$ \\\hline
\end{tabular}
\caption{Assignment of the $Z_2$ parity and $\xi_f$ factors in Eq.~(\ref{yukawa_int}) in each type of the Yukawa interactions.}\label{tab:z2parity_for_2hdm}
\end{table}

In this section, we define the Lagrangian of the 2HDM,
in which the Higgs sector is composed of two $SU(2)_L$ doublet scalar fields $H_1$ and $H_2$. 
To avoid the tree level FCNC, we impose a $Z_2$ symmetry in the Higgs sector which can be softly-broken in general. 
Under the $Z_2$ parity, 
four types of Yukawa interactions are defined depending on the assignment of $Z_2$ charge as listed in Table~\ref{tab:z2parity_for_2hdm}. 

The most general Higgs potential with the softly-broken $Z_2$ parity is given as
\begin{align}
V = &
m_{11}^2 |H_1|^2
+ m_{22}^2 |H_2|^2
- ( m_{12}^2 H_1^\dagger H_2 + h.c.) \nonumber\\
& + \frac{\l_1}{2} |H_1|^4
+ \frac{\l_2}{2} |H_2|^4
+ \l_3 |H_1|^2 |H_2|^2
+ \l_4 |H_1^\dagger H_2|^2
+ \left[ \frac{\l_5}{2} (H_1^\dagger H_2)^2 + h.c. \right]. \label{eq:higgspotential}
\end{align}
Throughout the paper, we consider the CP-conserving case of the Higgs sector for simplicity, 
so that the imaginary parts of $m_{12}^2$ and $\lambda_5$ are assumed to be zero. 
The Higgs fields are parametrized as
\begin{align}
H_i = \left[\begin{array}{c}
h_i^+ \\
\frac{1}{\sqrt{2}}(v_i + h_i - i a_i )
\end{array}\right],\quad (i=1,2), 
\end{align}
where $v_1$ and $v_2$ are the VEVs of the Higgs doublets which are related to the Fermi constant $G_F$ by 
$v^2\equiv v_1^2+v_2^2 =1/(\sqrt{2}G_F)$. 
The ratio of the two VEVs is parametrized by $\tan\beta=v_2/v_1$. 

The mass eigenstates of the scalar bosons are expressed by introducing the mixing angles $\alpha$ and $\beta$ as
\begin{align}
\left(\begin{array}{c}
h_1 \\ h_2
\end{array}\right) 
&=
\left(\begin{array}{cc}
\cos\a & -\sin\a \\
\sin\a & \cos\a
\end{array}\right)
\left(\begin{array}{c}
H \\ h^0
\end{array}\right),\\
\left(\begin{array}{c}
a_1 \\ a_2
\end{array}\right) 
&=
\left(\begin{array}{cc}
\cos\b & -\sin\b \\
\sin\b & \cos\b
\end{array}\right)
\left(\begin{array}{c}
G^0 \\ A
\end{array}\right),\\
\left(\begin{array}{c}
h_1^\pm \\ h_2^\pm
\end{array}\right) 
&=
\left(\begin{array}{cc}
\cos\b & -\sin\b \\
\sin\b & \cos\b
\end{array}\right)
\left(\begin{array}{c}
G^\pm \\ H^\pm
\end{array}\right), 
\end{align}
where $G^0$ and $G^\pm$ are the Nambu-Goldstone bosons which are absorbed 
by the $Z$ and $W$ bosons as the longitudinal component, respectively. 

The squared masses for the physical Higgs bosons are given by 
\begin{align}
m_{H^\pm}^2&=M^2-\frac{v^2}{2}(\lambda_4+\lambda_5), \notag \\
m_A^2&=M^2-v^2\lambda_5,   \notag \\
m_H^2&=\cos^2(\alpha-\beta)M_{11}^2+\sin^2(\alpha-\beta)M_{22}^2+\sin2(\alpha-\beta)M_{12}^2, \notag \\
m_h^2&=\sin^2(\alpha-\beta)M_{11}^2+\cos^2(\alpha-\beta)M_{22}^2-\sin2(\alpha-\beta)M_{12}^2,   \label{masssq}
\end{align}%
where $M^2=m_{12}^2/(\sin\beta \cos\beta)$ describes the breaking scale of the $Z_2$ symmetry, and $M_{ij}^2$ are 
given by 
\begin{align}
M_{11}^2&=v^2(\lambda_1\cos^4\beta+\lambda_2\sin^4\beta)+\frac{v^2}{2}\lambda_{345}\sin^22\beta,  \label{m11}  \\
M_{22}^2&=M^2+v^2(\lambda_1+\lambda_2-2\lambda_{345})\sin^2\beta\cos^2\beta, \label{m22}  \\
M_{12}^2&=\frac{v^2}{2}(\lambda_2\sin^2\beta-\lambda_1\cos^2\beta+\lambda_{345}\cos2\beta)\sin2\beta,  \label{m12}
\end{align}
where $\lambda_{345}=\lambda_3+\lambda_4+\lambda_5$. 
The mixing angle $\alpha$ is also  expressed in terms of $M_{ij}^2$ as
\begin{align}
\tan 2(\alpha-\beta)=\frac{2M_{12}^2}{M_{11}^2-M_{22}^2}. \label{tan2a}
\end{align} 
All the quartic coupling constants in the Higgs potential can be rewritten in terms of the physical parameters as    
\begin{align}
\lambda_1v^2 &= -M^2\tan^2\beta + (m_H^2\tan^2\beta + m_h^2)s^2_{\beta-\alpha} 
             +(m_H^2 + m_h^2\tan^2\beta)c^2_{\beta-\alpha} \notag\\
             &\quad\quad+2(m_H^2-m_h^2)\tan\beta s_{\beta-\alpha}c_{\beta-\alpha}, \notag\\
\lambda_2v^2 &=  -M^2\cot^2\beta +(m_H^2\cot^2\beta + m_h^2)s^2_{\beta-\alpha}
             +(m_H^2 + m_h^2\cot^2\beta)c^2_{\beta-\alpha} \notag\\
             &\quad\quad-2(m_H^2-m_h^2)\cot\beta s_{\beta-\alpha}c_{\beta-\alpha} , \notag\\
\lambda_3v^2 &=2m_{H^\pm}^2-M^2  +(m_h^2- m_H^2)[s^2_{\beta-\alpha}-c^2_{\beta-\alpha} -(\tan\beta-\cot\beta)s_{\beta-\alpha}c_{\beta-\alpha}], \notag\\
\lambda_4v^2&=M^2+m_A^2-2m_{H^\pm}^2,\notag\\
\lambda_5v^2&=M^2-m_A^2, \label{physical}
\end{align} 
where $s_{\beta-\alpha}=\sin(\beta-\alpha)$ and $c_{\beta-\alpha}=\cos(\beta-\alpha)$. 

A size of some combinations of $\lambda$'s in the Higgs potential is constrained 
by taking into account perturbative unitarity~\cite{Uni-2hdm1,Uni-2hdm2,Uni-2hdm3,Uni-2hdm4}
and vacuum stability~\cite{VS_THDM,VS_THDM2}. 
Through Eq.~(\ref{physical}), such a constraint can be translated into a bound on the physical parameters; \textit{e.g.}, the masses of the scalar bosons. 
First, the condition for vacuum stability; \textit{i.e.}, the requirement for bounded from below in any direction of the Higgs potential with large scalar fields, 
is given by~\cite{VS_THDM,VS_THDM2} 
\begin{align}
\lambda_1>0, \quad \lambda_2>0,\quad \sqrt{\lambda_1\lambda_2}+\lambda_3+\text{MIN}(0,\lambda_4+\lambda_5,\lambda_4-\lambda_5)>0.
\end{align}
Second, the perturbative unitarity bound is obtained by requiring that 
all the eigenvalues of the $s$-wave amplitude matrix $a_{i,\pm}^0$ for the elastic scatterings of two body boson states
are satisfied as 
\begin{align}
|a_{i,\pm}^0|\leq\frac{1}{2}. \label{pv1}
\end{align}
All the independent eigenvalues $a_{i,\pm}^0$ were derived in Refs.~\cite{Uni-2hdm2,Uni-2hdm3,Uni-2hdm4} as
\begin{align}
a_{1,\pm}^0 &=  \frac{1}{32\pi}
\left[3(\lambda_1+\lambda_2)\pm\sqrt{9(\lambda_1-\lambda_2)^2+4(2\lambda_3+\lambda_4)^2}\right],\\
a_{2,\pm}^0 &=
\frac{1}{32\pi}\left[(\lambda_1+\lambda_2)\pm\sqrt{(\lambda_1-\lambda_2)^2+4\lambda_4^2}\right],\\
a_{3,\pm}^0 &= \frac{1}{32\pi}\left[(\lambda_1+\lambda_2)\pm\sqrt{(\lambda_1-\lambda_2)^2+4\lambda_5^2}
\right],\\
a_{4,\pm}^0 &= \frac{1}{16\pi}(\lambda_3+2\lambda_4\pm 3\lambda_5),\\
a_{5,\pm}^0 &= \frac{1}{16\pi}(\lambda_3\pm\lambda_4),\\
a_{6,\pm}^0 &= \frac{1}{16\pi}(\lambda_3\pm\lambda_5).  \label{pv2}
\end{align}

The Yukawa interaction terms are given by 
\begin{align}
{\cal L}_{\rm Yukawa} =& -y_u \tilde H_u^T Q_L u_R^c - y_d H_d^\dagger Q_L d_R^c - y_\ell H_\ell^\dagger L_L e_R^c + h.c.,\label{yukawa}
\end{align}
where $\tilde H_u = i\t^2 H_u$.
In Eq.~(\ref{yukawa}), $H_u$, $H_d$ and $H_\ell$ are either $H_1$ or $H_2$ depending on the type of Yukawa interaction. 
In the mass eigenstates of the scalar bosons, the interaction terms are expressed as
\begin{align}
{\cal L}_{\rm Yukawa} =
 & -\sum_{f=u,d,\ell}\frac{m_f}{v}\left(\xi_f^h \, h \bar ff + \xi_f^H \, H \bar ff - 2iT_f^3 \xi_f \, A \bar f\gamma_5 f \right) \nonumber\\
 & + \left[ \sqrt{2}V_{ud} H^+ \bar u \left( \frac{m_u\xi_u}{v} P_L 
- \frac{m_d\xi_d}{v} P_R\right) d  -\frac{\sqrt{2} m_\ell \xi_\ell}{v} H^+ \bar \n P_R \ell + h.c.\right], \label{yukawa_int}
\end{align}
where $T_f^3=+1/2~(-1/2)$ for $f=u$ ($d,\ell$), and $V_{ff'}$ is the Cabibbo-Kobayashi-Maskawa matrix element. 
The $\xi_f^h$ and $\xi_f^H$ factors are defined by 
\begin{align}
\xi_f^h = s_{\beta-\alpha}+\xi_f c_{\beta-\alpha},\quad
\xi_f^H = c_{\beta-\alpha}-\xi_f s_{\beta-\alpha}. \label{xih}
\end{align}
The $\xi_f$ factors in Eqs.~(\ref{yukawa_int}) and (\ref{xih}) are given in Table~\ref{tab:z2parity_for_2hdm}.

From the kinetic terms of the scalar fields, 
the ratios of the coupling constant among the CP-even scalars and gauge bosons are extracted as
\begin{align}
\frac{g_{hVV}}{g_{hVV, {\rm SM}}} = s_{\b-\a},\qquad
\frac{g_{HVV}}{g_{hVV, {\rm SM}}} = c_{\b-\a}.\qquad(V=W,Z) \label{hvv}
\end{align}
As it is seen in Eqs.~(\ref{yukawa_int}), (\ref{xih}) and (\ref{hvv}),  
in the limit of $\sin(\b-a)\to 1$, 
both $hf\bar{f}$ and $hVV$ couplings become the same as those in the SM, so that we can call this limit as the SM-like limit.

\section{Constraints on the Type-X 2HDM}\label{sec:constraints}

\begin{figure}[t]
\centering
\includegraphics[width=0.3\hsize]{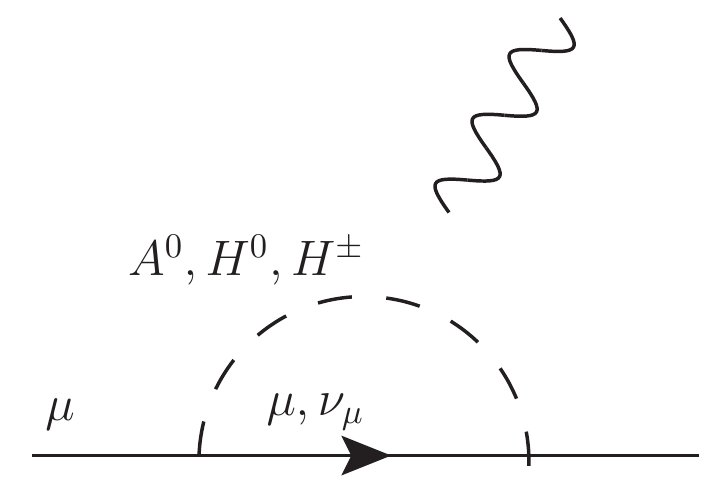}
\qquad\qquad
\includegraphics[width=0.3\hsize]{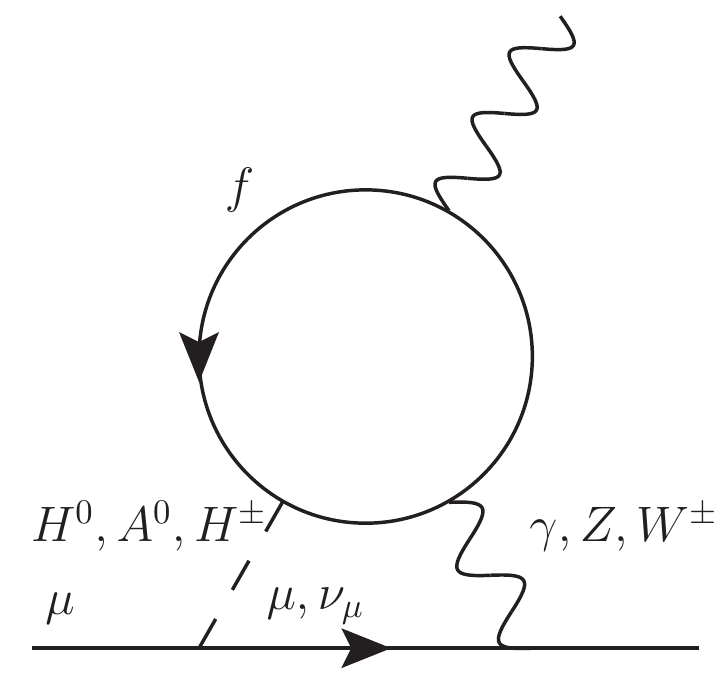}
\caption{
One-loop (left) and two-loop Barr-Zee (right) diagrams which give corrections to the muon $g-2$.
}\label{fig:amu}
\end{figure}
In the 2HDMs, 
the one-loop diagrams and the two-loop Barr-Zee type diagrams shown in Fig.~\ref{fig:amu} 
give dominant contributions to the muon $g-2$.
It has been known that 
the Barr-Zee type diagrams give a sizable positive contribution to $a_\mu$ in the case of 
a large $A\ell^+\ell^-$ coupling and a small $m_A$ as pointed it out in Refs.~\cite{Chang:2000ii,Kingman1}. 
In the Type-X 2HDMs, a large $A\ell^+\ell^-$ can be realized by taking 
$\tan\beta \gg 1$ since $\xi_\ell=-\tan\beta$ as shown in Table~\ref{tab:z2parity_for_2hdm}. 
Typically, when $\tan\beta \gtrsim 40$ and $m_A=\mathcal{O}(10\text{-}100)$ GeV, 
the muon $g-2$ anomaly can be explained in the Type-X 2HDM~\cite{Chun}.
In this section, we focus on the Type-X 2HDM with the large $\tan\beta$ and small $m_A$ scenario to explain the $g-2$ anomaly, and 
we discuss important experimental constraints in this situation.

\subsection{Direct searches for the extra Higgs bosons}

There has been no signal of the extra Higgs bosons at any collider experiments.
This gives lower limits on the masses of the extra Higgs bosons depending on the magnitude of couplings with SM particles. 
We first summarize the current bounds from the LEP II experiment, and we
also review those from the LHC Run-I.

\subsubsection{LEP II}\label{sec:LEP2}

There are constraints on the masses of the extra Higgs bosons from the direct production at the LEP II experiment 
with the maximal collision energy to be about 200 GeV. 
From the $H^\pm$ pair production process $e^+ e^- \to \g^* / Z^* \to H^+ H^-$ 
the lower bound was obtained by $m_{H^\pm} > 93.5~{\rm GeV}$ at 95~\%~C.L.~\cite{Abbiendi:2013hk} under the assumption of ${\rm Br}(H^+ \to\t^+ \n_\t) =1$ 
which is realized by $\tan\beta\gtrsim 2$, $\sin(\beta-\alpha)\simeq 1$ in the Type-X 2HDM.

From the pair production of the neutral Higgs bosons $e^+ e^- \to Z^* \to AH$, 
the lower bound for the sum of $m_A$ and $m_H$ is given to be about 190-195 GeV for $m_A > 40~{\rm GeV}$~\cite{Schael:2006cr} 
under the assumption of ${\rm Br}(H \to \t^+\t^-) = {\rm Br}(A \to \t^+\t^-) =1$ which is realized by $\tan\beta \gtrsim 3$ and $\sin(\beta-\alpha)\simeq 1$ in the 
Type-X 2HDM. 

The searches for $A$ and $H$ from the bremsstrahlung process $e^+ e^- \to \t\t A/H $ have also been 
performed for the range of $m_{A/H}<50~\text{GeV}$. 
This process gives an upper bound on $\tan\beta$ for a fixed value of $m_{A/H}$. 
For example, 
$\tan\beta>128.1~(120.1)$ and $\tan\beta>44.8~(40.0)$ are respectively excluded at 95\% C.L.~for $m_A$ ($m_H$) to be 30 GeV and 
15 GeV~\cite{Abdallah:2004wy} with the case of Br($A\to \tau\tau$)$=$Br($H\to \tau\tau$)=1. 

We note that the branching fractions for the extra Higgs bosons into a fermion pair can be reduced when 
there is a non-zero mass splitting among them. 
For example, $H^\pm \to AW^{(*)}$ and $H\to AZ^{(*)}$ open in the case of $m_{H^\pm}> m_A$ and $m_H>m_A$, respectively. 
There also happen the inverse processes like $A \to H^\pm W^{\mp (*)}$ and $A\to HZ^{(*)}$ as long as they are kinematically allowed. 
In such decay modes associated with a gauge boson,  
the bounds on masses on the extra Higgs bosons can be weaker than those given in the above. 

\subsubsection{LHC Run-I}\label{sec:LHC1}

At the LHC, extra Higgs boson searches have been performed in various channels. 
In the most of channels, an enhancement of the Yukawa couplings of
the extra Higgs bosons becomes important to 
obtain a bound on their masses or coupling constants. 
However, in the Type-X 2HDM, the couplings of the neutral extra Higgs bosons
to the quarks are suppressed by $\cot\beta$. 
Thus, the processes such as $gg\to A/H \to \tau\tau$ 
and $gg\to b\bar{b}A/b\bar{b}H \to b\bar{b} \tau\tau$ 
do not set a limit on the masses in a large $\tan\beta$ case.

Similar to the neutral Higgs boson productions, the cross section of the $H^\pm$ production such as $gb\to H^\pm t$ 
is also suppressed by $\cot^2\beta$ in the Type-X 2HDM. 
If $m_{H^\pm}+m_b<m_t$, the top decay $t\to H^\pm b$ can be used to constrain $m_{H^\pm}$. 
From the process $pp\to t\bar{t}\to b\bar{b}H^\pm W^\mp$ with $H^\pm \to \tau^\pm \nu$,  
the upper limit on BR$(t\to H^\pm b)\times$BR$(H^\pm \to \tau^\pm\nu)$ has been driven 
to be between 0.23\% and 1.3\% at 95\% C.L.~for $m_{H^\pm}$ in the range of 80 GeV to 160 GeV~\cite{ATLAS_H+}. 
This gives the bounds, for example, $\tan\beta \lesssim 6$ and 15 for $m_{H^\pm}=100$ and 150 GeV at 95\% C.L.~in the Type-X 2HDM using 0.23\%
of the product of the branching fractions.

Apart from the production processes via Yukawa couplings, 
one must take care of the $h\to AA$ decay in the case of $m_A<m_h/2$. 
In the Type-X 2HDM, this typically gives the four $\tau$ final state, because the $A\to \tau\tau$ decay can be the main decay mode as explained in Sec.~\ref{sec:LEP2}.
In Ref.~\cite{Curtin:2013fra}, the upper bound on ${\rm Br}(h\to AA\to 4\t)$ is given to be about $0.2$ for $m_A > 30~{\rm GeV}$
and $0.2$-$0.5$ for $15<m_A<30~{\rm GeV}$.
In the 2HDMs, the branching fraction is determined by the dimensionless $hAA$ coupling $\lambda_{hAA}$
defined as the coefficient of the $hAA$ vertex in the Lagrangian; \textit{i.e.},  ${\cal L} = v \l_{hAA}\, h AA+\cdots$ which is given by 
\begin{align}
\lambda_{hAA}&=\frac{1}{2v^2}\left[(2M^2-2m_A^2-m_h^2)s_{\beta-\alpha}+(M^2-m_h^2)(\cot\beta-\tan\beta) c_{\beta-\alpha}  \right]. \label{lam_haa}
\end{align}
The partial decay width of $h\to AA$ is then expressed by 
\begin{align}
\G_{h\to AA}
~=~ \frac{\l_{hAA}^2 v^2}{8\pi m_h}\sqrt{1-\frac{4m_A^2}{m_h^2}}
~\simeq~ \G_{\rm SM} \times \left( \frac{\l_{hAA}}{0.015} \right)^2 \sqrt{1-\frac{4m_A^2}{m_h^2}},  
\end{align}
where $\Gamma_{\text{SM}}=4.41$ MeV is the total decay width of the SM Higgs boson for $m_h=125~\GEV$~\cite{Djouadi:1997yw}.
Therefore, to satisfy Br($h\to AA$)$<0.2$, $\l_{hAA} \lsim 6.7\times 10^{-3}$ is required. 
We can simply take $\lambda_{hAA}=0$ by setting an appropriate value of $\beta-\alpha$ from Eq.~(\ref{lam_haa}) as 
\begin{align}
\tan(\beta-\alpha) = \frac{M^2-m_h^2}{2M^2-2m_A^2-m_h^2}(\tan\beta-\cot\beta). 
\end{align}
In the case of $\tan\beta\gg 1$, $m_h^2/m_{H^\pm}^2 \ll 1$ and $m_A^2/m_{H^\pm}^2 \ll 1$, we obtain 
\begin{align}
\sin(\beta-\alpha)&\simeq 1-\frac{2}{\tan^2\beta}\left(1+\frac{m_h^2}{m_{H^\pm}^2}-\frac{2m_A^2}{m_{H^\pm}^2}  \right), \label{sin_app}  \\
\cos(\beta-\alpha)&\simeq \frac{2}{\tan\beta}\left(1+\frac{m_h^2}{2m_{H^\pm}^2}-\frac{m_A^2}{m_{H^\pm}^2}  \right).  \label{cos_app}
\end{align}
From the above expressions, we find that the SM-like behavior of $h$ is realized by taking $\tan\beta \gg 1$, because of $\sin(\beta-\alpha)\simeq 1$. 


\subsection{Electroweak precision observables}

The extra Higgs bosons can modify the electroweak precision observables from the SM prediction via the loop effects.
Such an effect can be used as an indirect search for the extra Higgs bosons and also used to constrain parameter space in the 2HDM. 
In this subsection, we discuss the constraints from the oblique parameters and the $Z$ boson decay. 

\subsubsection{Oblique parameters}

The electroweak oblique $S$, $T$ and $U$ parameters are introduced by Peskin and Takeuchi~\cite{Peskin:1991sw} which 
parametrize new physics effects on the gauge boson two point functions. 
These parameters are calculated in 2HDMs in Refs.~\cite{Toussaint:1978zm, Bertolini:1985ia, Pomarol:1993mu, Peskin:2001rw, Gerard:2007kn, Grimus:2008nb, Kanemura:2011sj}.
In the SM-like limit $\sin(\b-\a) \to 1$, 
these parameters are given to be the same formulae 
as those given in the inert doublet model \cite{Barbieri:2006dq}.  
For the case of $m_A \ll m_Z \ll m_{H^\pm}\simeq m_H$, the contribution
from the additional scalar bosons is given by 
\begin{align}
\D S &~\simeq~ -\frac{5}{72\pi} ~\simeq~ 0.022, \\
\D T &~\simeq~ \frac{m_H(m_{H^\pm}-m_H)}{32\pi^2 \alpha_{\text{em}} v^2} ~\simeq~ 0.013 \times\left( \frac{m_H}{200~{\rm GeV}} \right)\times\left(\frac{m_{H^\pm}-m_H}{10~{\rm GeV}} \right).
\end{align}
We also find that $\Delta U$ is the same order as
$\Delta S$ in our setup for large $\Delta T$ regime.
If we take $m_H = m_{H^\pm}$ and $\sin(\beta-\alpha)=1$,
the Higgs potential respects the custodial $SU(2)_V$ symmetry~\cite{Pomarol:1993mu, Gerard:2007kn},
which makes $\D T=0$. 
The $S$ and $T$ parameters driven by the Gfitter group \cite{Baak:2014ora} are
\begin{align}
\D S=0.05\pm 0.11,\qquad \D T=0.09\pm 0.13, 
\end{align}
with the reference values of $m_h=125$ GeV and $m_t=173$ GeV. 
The prediction of $\Delta S$ parameter is inside the $1\sigma$ error of the measured value,
and the $T$ parameter constrains on the mass splitting $|m_H-m_{H^\pm}| ={\cal O}(10)~{\rm GeV}$.
Hence we take $m_{H}= m_{H^{\pm}}$ to avoid the constraint on
the oblique parameters in the following analysis.

\subsubsection{$Z$ boson decay}

The property of the $Z$ boson such as the mass, the total width, and the decay branching ratios  
were precisely measured at the LEP experiment. 
If new physics particles modify such a precisely measured quantity, 
their masses and/or couplings are severely constrained. 

In our scenario, the $Z\tau^+\tau^-$ vertex can be significantly deviated from the SM prediction by 
loop effects of the extra Higgs bosons, because 
they strongly interact with charged leptons in the large $\tan\beta$ case.  
In order to discuss how the modified vertex affects the observables, we define 
the effective $Zf\bar{f}$ vertex as 
\begin{align}
\mathcal{L} = g_Z\bar{f}\gamma^\mu(\hat{v}_f-\hat{a}_f\gamma_5)fZ_\mu, 
\end{align}
where $g_Z=g/\cos\theta_W$ and $\theta_W$ being the weak mixing angle. 
Although there are several definitions for $\sin^2\theta_W$, we 
here use the on-shell definition~\cite{sirlin} of it which is determined by using $W$ and $Z$ boson masses, \textit{i.e.}, $\sin^2\theta_W=1-m_W^2/m_Z^2$. 
The effective vector coupling $\hat{v}_f$ and axial vector coupling $\hat{a}_f$ can be separately written by the contributions from the tree level and from the one-loop level as
\begin{align}
\hat{v}_f &= v_f + v_f^{\text{loop}}, \quad
\hat{a}_f = a_f + a_f^{\text{loop}}, 
\end{align}
where the tree level contributions are expressed as
\begin{align}
v_f = \frac{1}{2}T_f^3 -\sin^2\theta_W Q_f, \quad a_f =  \frac{1}{2}T_f^3, \label{vector_coupling}
\end{align}
with $Q_f$ being the electric charge of $f$. 
The loop contributions $v_f^{\text{loop}}$ and $a_f^{\text{loop}}$ 
are composed of the counter term and the one particle irreducible (1PI) $Zf\bar{f}$ vertex correction diagram:
\begin{align}
v_f^{\text{loop}} &= \delta v_f + v_f^{\text{1PI}}, \quad 
a_f^{\text{loop}} = \delta a_f + a_f^{\text{1PI}}. 
\end{align} 
After imposing the on-shell renormalization conditions, 
the counter term contribution is expressed by~\cite{Hollik_EW} 
\begin{align}
\delta v_f&=
-a_f\Pi_{ff,A}^{\text{1PI}}(m_f^2)-v_f\left[\Pi_{ff,V}^{\text{1PI}}(m_f^2)-2m_f^2\frac{d}{dp^2}[\Pi_{ff,V}^{\text{1PI}}(p^2)+\Pi_{ff,S}^{\text{1PI}}(p^2)]_{p^2=m_f^2}\right],\label{lamhatv}\\
\delta a_f & = 
-a_f\left[\Pi_{ff,V}^{\text{1PI}}(m_f^2)+2m_f^2\frac{d}{dp^2}[\Pi_{ff,V}^{\text{1PI}}(p^2)+\Pi_{ff,S}^{\text{1PI}}(p^2)]_{p^2=m_f^2}\right]
-v_f\Pi_{ff,A}^{\text{1PI}}(m_f^2), 
\end{align}
where $\Pi_{ff,X}^{\text{1PI}}$ are the 1PI diagram contributions to the fermion two point functions defined as 
\begin{align}
\Pi_{ff}^{\text{1PI}}(p^2) = 
p\hspace{-2mm}/\Pi_{ff,V}^{\text{1PI}}(p^2) - p\hspace{-2mm}/\gamma_5\Pi_{ff,A}^{\text{1PI}}(p^2)  
+m_f\Pi_{ff,S}^{\text{1PI}}(p^2). 
\end{align}
In the SM-like limit $\sin(\beta-\alpha)=1$, 
the deviation in $v_\tau^{\text{loop}}$ and $a_\tau^{\text{loop}}$ purely comes from the 
extra Higgs boson loop diagrams. In this case for $f=\tau$, we obtain 
\begin{align}
&\Delta v_\tau^{\text{loop}} = v_\tau^{\text{loop}}-v_\tau^{\text{loop}}(\text{SM})\notag\\
&\simeq \frac{1}{16\pi^2}\left(\frac{m_\tau}{v}\xi_\ell\right)^2
\Big[v_\tau F_1(m_H)+v_\tau F_1(m_A)-2a_\tau F_1(m_{H^\pm})+(v_\tau+a_\tau)F_2(m_{H^\pm},m_{H^\pm}) \Big], \\
 &\Delta a_\ell^{\text{loop}}= a_\tau^{\text{loop}}-a_\tau^{\text{loop}}(\text{SM})\notag\\
&\simeq -\frac{1}{16\pi^2}\left(\frac{m_\tau}{v}\xi_\ell\right)^2
\Big[a_\tau F_1(m_H)+a_\tau F_1(m_A)-2a_\tau F_1(m_{H^\pm})\notag\\
&\quad\quad\quad\quad\quad\quad\quad\quad\quad+(v_\tau+a_\tau) F_2(m_{H^\pm},m_{H^\pm})
-4a_\tau F_2(m_{H},m_A) \Big], 
\end{align}
where the loop functions are given as
\begin{align}
F_1(m)&=
\ln \frac{m^2}{m_Z^2}+\frac{5}{4}-i\pi+\frac{1}{2}\int_0^1 dx \int_0^1 dy\frac{m_Z^2xy(2y-1)+m^2(2+y-yx)}{m_Z^2x(1-y)-m^2(1-x)} , \\
F_2(m_1,m_2)&=
\frac{1}{4}\left(\ln \frac{m_2^2}{m_1^2}+1\right) - \int_0^1 dx \int_0^1 dy\frac{y^2}{2}\frac{m_2^2-x[m_1^2-m_Z^2(1-2y)]}{m_2^2(1-y)+xy[m_1^2-m_Z^2(1-y)]}. 
\end{align}
In the above expressions, we neglect the mass of the tau lepton in the loop functions.  
We note that the $F_2(m_1,m_2)$ function is invariant under the interchange of $m_1\leftrightarrow m_2$, so that 
$\Delta v_\tau^{\text{loop}}$ and $\Delta a_\tau^{\text{loop}}$ does not change the value under $m_H\leftrightarrow m_A$. 

Let us apply the modified $Z\tau^+\tau^-$ vertex to 
the leptonic partial decay width of the $Z$ boson $Z\to \ell\ell$:
\begin{align}
\Gamma(Z\to \ell\ell) = g_Z^2\frac{m_Z}{12\pi}(\hat{v}_\ell^2+\hat{a}_\ell^2). 
\end{align}
We define the ratio of the partial width of $Z\to \tau\tau$ to that of $Z\to ee$ as
\begin{align}
R_{\tau/e}  &\equiv \Gamma(Z\to \tau\tau)/\Gamma(Z\to ee). 
\end{align}
The deviation in the ratio from the SM predictions are then given by 
\begin{align}
\Delta R_{\tau/e} &\equiv   R_{\tau/e} -  R_{\tau/e}^{\text{SM}}
\simeq \frac{1}{\Gamma(Z\to \ell\ell)_{\text{SM}}}\frac{g_Z^2}{6\pi}m_Z^{}\left[v_\ell \text{Re}\Delta v_\ell^{\text{loop}} + a_\ell \text{Re}\Delta a_\ell^{\text{loop}}\right]. 
\end{align} 
The SM prediction is given by 
\begin{align}
\Gamma(Z\to \ell\ell)_{\text{SM}} = 83.995\pm 0.010~\text{MeV}. 
\end{align}
The measured values of the leptonic decay width and $R_{\tau/e}$ are given by \cite{PDG2014}
\begin{align}
&\Gamma(Z\to \ell\ell)_{\text {exp}} = 83.984\pm 0.086 \text{ MeV}, \quad
R_{\tau/e}^{\text {exp}} = 1.0019 \pm 0.0032. 
\end{align}
We find that $\tan\beta > 50~(70)$ is excluded for $m_A=10$~(50) GeV when
$m_{H} = m_{H^{\pm}}=300$~GeV. The bound becomes weaker for the larger
$m_{H^{\pm}}$. We will combine the constraint from the $Z\to \ell\ell$ decay 
with the muon $g-2$ result in Sec.~\ref{sec:muon_g-2}.

\subsection{Flavor experiments}\label{sec:flavor}

Effects of the extra Higgs bosons can appear in various observables measured at flavor experiments. 
Therefore, similar to the electroweak precision measurements, 
flavor measurements can be used to constrain the parameter space in the 2HDM. 
In this subsection, we discuss $B_s \to \m^+ \m^-$ and the leptonic decay of $\t$.

\subsubsection{$B_s \to \mu^+\mu^-$}\label{sec:bsmm}

It was pointed out in Ref.~\cite{Logan:2000iv} that
the branching fraction of the $B_s \to \ell\ell$ process in the
Type-II 2HDM is enhanced by the factor $\tan^4\beta$ which comes from 
the box type and the penguin type diagrams with extra Higgs boson mediation. 
The origin of the $\tan^4 \beta$ dependence
is that both the charged lepton and down-type quark couplings to the extra Higgs 
bosons are proportional to $\tan\beta$ in the Type-II 2HDM. 
On the other hand, in the Type-X 2HDM, the lepton couplings are enhanced by $\tan\b$, while the 
quark couplings are suppressed by $\cot\b$. 
Thus, the $\tan^4\beta$ dependence does not appear in the branching fraction of $B_s \to \ell\ell$, and 
the additional leading contribution is almost independent on $\tan\b$ for large $\tan\b$.
Although the deviation from the SM becomes mild as compared to the case of Type-II, 
we check ${\rm Br}(B_s \to \mu^+ \mu^-)$ because the light CP-odd Higgs boson $A$ could give a sizable contribution, which 
is required to explain muon $g-2$ anomaly. 

For the calculation of $B_s\to\mu^+\mu^-$ in the 2HDM, we use the formulae given in Ref.~\cite{Li:2014fea}.
The observed branching fraction at the LHC is ${\rm Br}(B_s \to \mu^+ \mu^-) = (2.9 \pm 0.7) \times 10^{-9}$ \cite{CMSandLHCbCollaborations:2013pla}.
We show the constraint from $B_s \to \m^+ \m^-$ on the parameter space of 2HDM in Sec.~\ref{sec:muon_g-2}.

\subsubsection{Leptonic $\tau$ decay at the tree level}\label{sec:taudecay}

\begin{figure}[p]
\centering
\includegraphics[width=0.48\hsize]{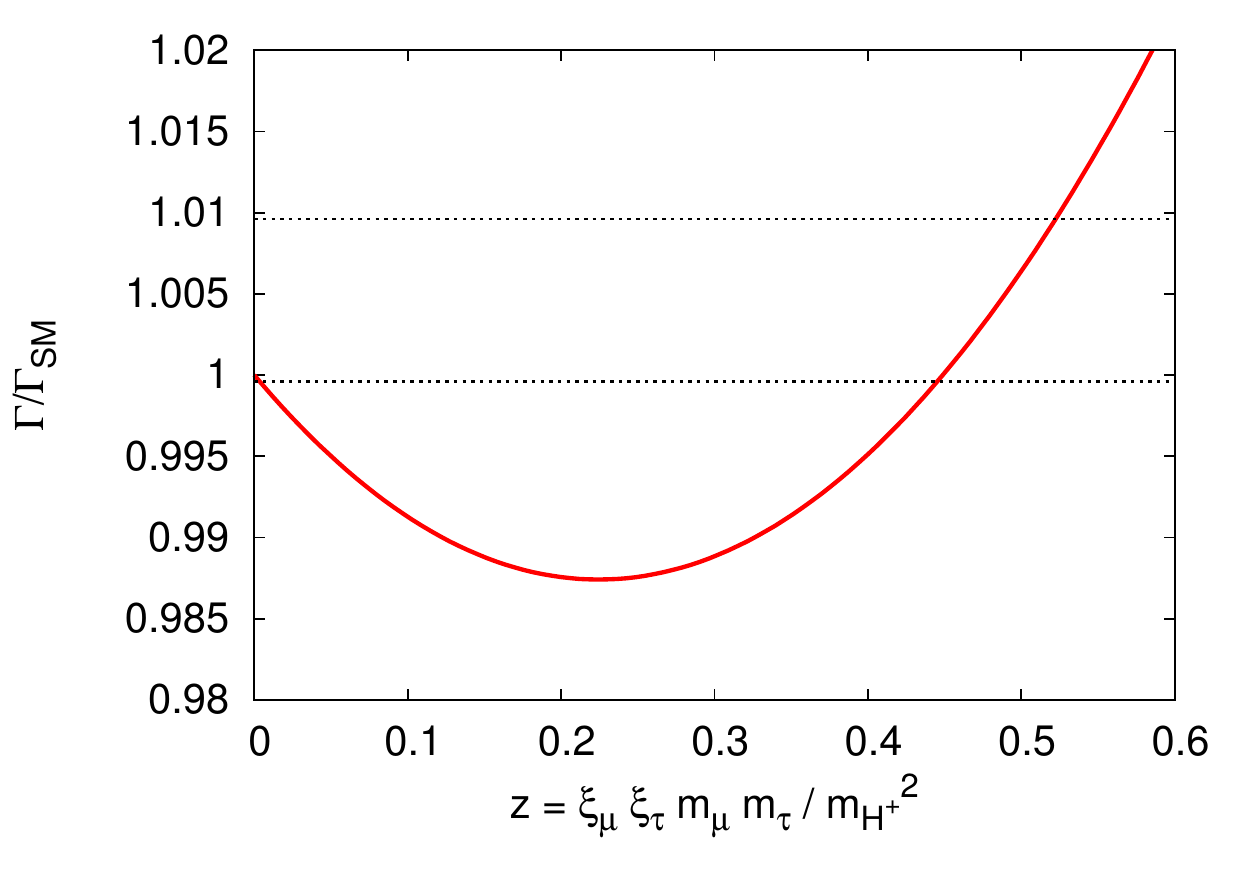} 
\includegraphics[width=0.48\hsize]{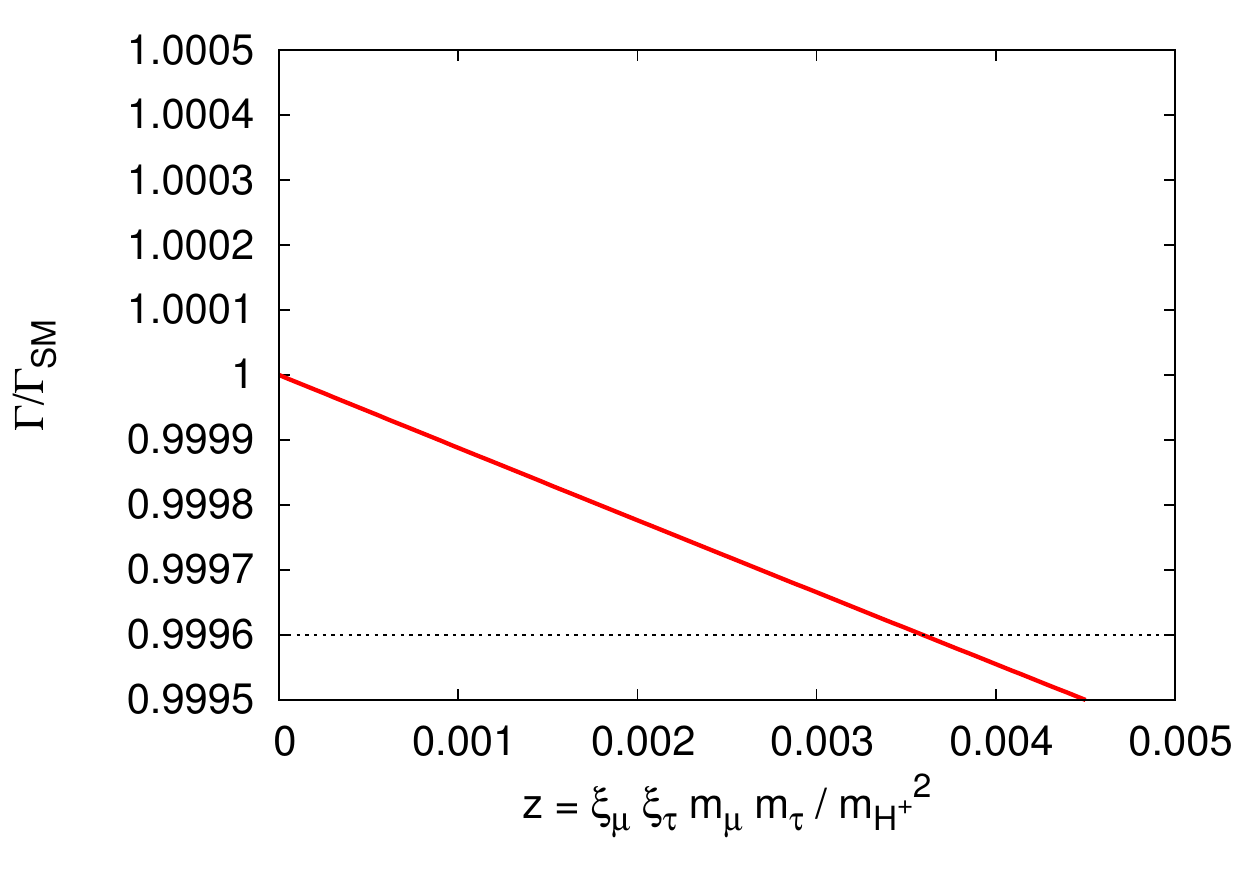} \\
\includegraphics[width=0.48\hsize]{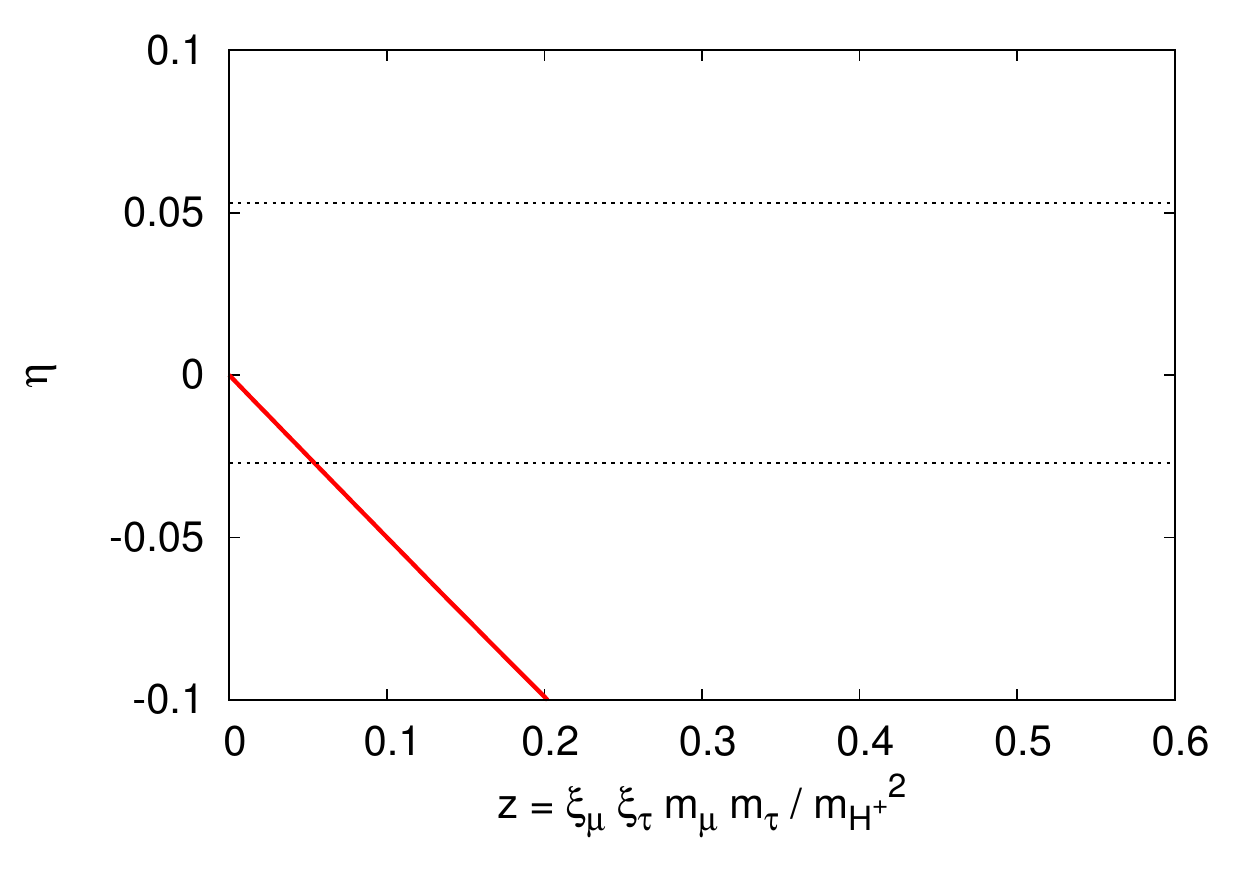} 
\includegraphics[width=0.48\hsize]{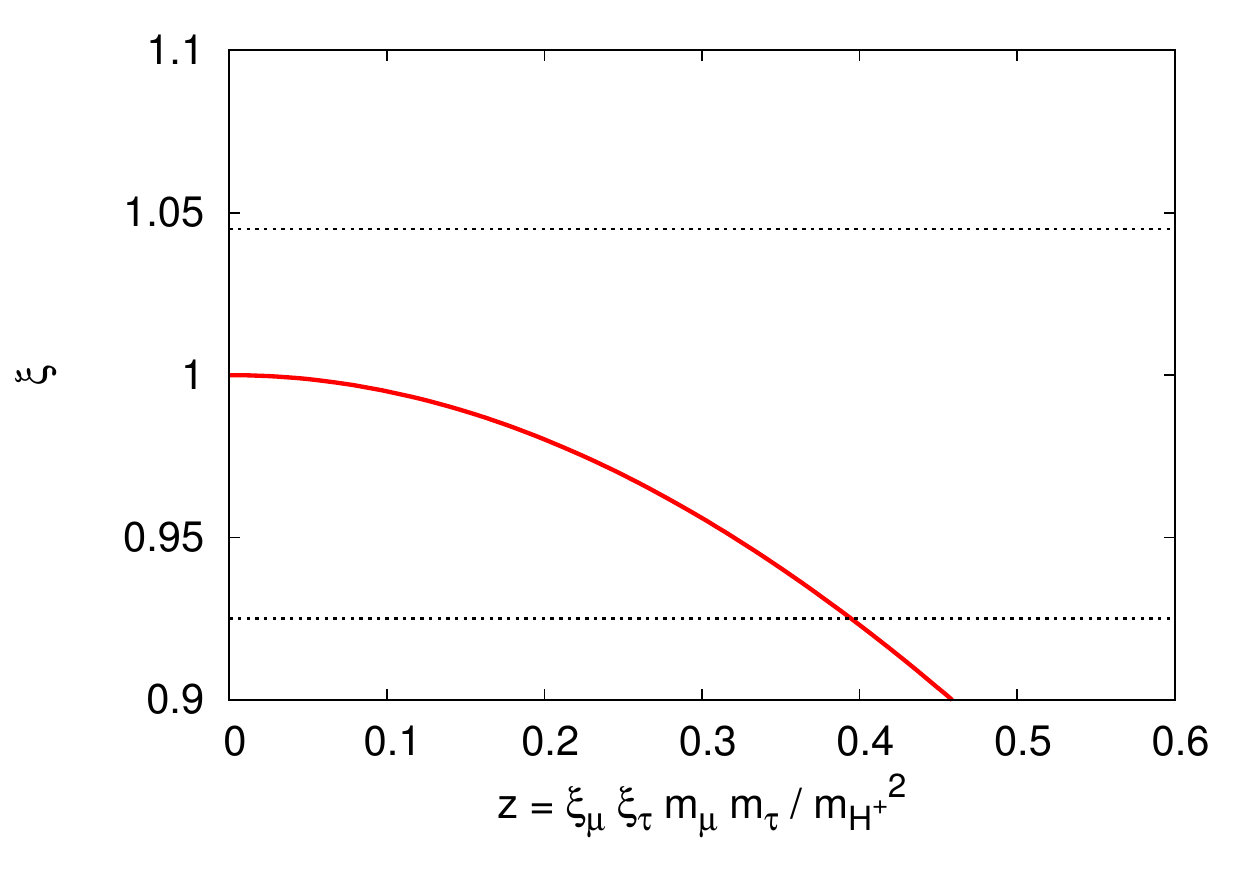}
\caption{
The ratio of the decay rate (upper left and upper right) and the Michel parameters $\eta$ (lower left) and $\xi$ (lower right)
as a function of $z$ in the leptonic tau decay.
In each panel, outside regions from the dotted lines are excluded at the 95~\% C.L.
}\label{fig:taudecay}
\end{figure}

In the SM, the leptonic $\tau$ decay is caused by the $W$ boson exchange diagram at tree level. 
In the 2HDM, the $H^\pm$ mediated diagram also contributes to the leptonic $\tau$ decay.  
The effect of $H^\pm$ contribution on the partial decay width of $\tau$ was calculated in Refs.~\cite{Krawczyk:2004na, Aoki:2009ha}, 
and that on the Michel parameters, which is defined just below, in Ref.~\cite{Logan:2009uf}.

The differential decay rate of $\tau \to \mu \nu_\mu \nu_\t$ is given in terms of 
the Michel parameters ($\rho,\eta,\d$ and $\xi$) and $\hat G_{\m\t}$ defined in Eqs.~(\ref{ghat}) and (\ref{michel})
as \cite{Pich:2013lsa}
\begin{align}
\frac{d^2 \G}{dxd\cos\theta} = \frac{m_\m \omega^4}{2\pi^3} \hat G_{\m\t}^2 \sqrt{x^2-x_0^2} \left( F(x) - \frac{\xi}{3}{\cal P}_\t \cos\theta \sqrt{x^2-x_0^2}A(x) \right),
\end{align}
where $\omega \equiv (m_\t^2+m_\m^2)/2m_\t$, $x \equiv E_\m / \omega$ and $x_0 \equiv m_\m / \omega$ with $E_\mu$ being the muon energy.
${\cal P}_{\tau}$ is the polarization of the tau, and
$\theta$ is the angle between the polarization and the momentum
direction of the muon.
The functions $F(x)$ and $A(x)$ are defined as
\begin{align}
F(x) &= x(1-x) + \frac{2}{9}\rho (4x^2-3x-x_0^2) + \eta x_0 (1-x),\\
A(x) &= 1-x + \frac{2}{3}\d (4x-4+\sqrt{1-x_0^2}).
\end{align}
By using $z \equiv m_\m m_\t \tan^2\b / m_{H^+}^2$,
we find\footnote{We find these expressions are 
inconsistent with Ref.~\cite{Logan:2009uf}}
\begin{align}
\hat G_{\m\t} = G_F \left( 1+\frac{z^2}{4} \right)^{1/2}, \label{ghat}\\
\rho = \frac{3}{4}, \qquad
\eta = -\frac{2z}{4 + z^2 }, \qquad
\delta = \frac{3}{4}, \qquad
\xi = \frac{4-z^2}{4+z^2}.  \label{michel}
\end{align}
We see that $\rho$ and $\d$ are equal to the SM values at the tree level.
The observed Michel parameters of the $\t$ decay are $\eta=0.013\pm 0.020$ and $\xi=0.985\pm 0.030$ \cite{PDG2014}.
The ratio of the decay rate in the 2HDM to that in the SM prediction is given as \cite{Krawczyk:2004na, Aoki:2009ha}
\begin{align}
\left(
\frac{G_{\mu \tau}}{G_F}
\right)^2
\equiv
\frac{\G(\t\to\m\n\n)_{\rm 2HDM}}{\G(\t\to\m\n\n)_{\rm SM}} =
1 - 2z \frac{m_\m}{m_\t} \frac{g(m_\m^2/m_\t^2)}{f(m_\m^2/m_\t^2)} + \frac{z^2}{4}, 
\label{taudecay}
\end{align}
where the phase functions $f(x)$ and $g(x)$ are given by $f(x) = 1-8x-12x^2\log x+ 8x^3-x^4$ and $g(x) = 1+9x-9x^2-x^3 + 6x(1+x)\log x$.  
To find a constraint on Eq.~(\ref{taudecay}), we can use the constraint
on the flavor universality. In the similar manner to Eq.~(\ref{taudecay}), we introduce $G_{e\mu}$ and $G_{e \tau}$.
Since $m_{e}, m_{\m} \ll m_{\tau}$,
the corresponding terms to
the rightest term in Eq.~(\ref{taudecay}) for $G_{e\mu}$ and $G_{e
\tau}$ are 1, and thus $G_{e\mu} = G_{e\tau} =G_F$ in 2HDM.
There are constraints on the lepton universality given by HFAG
group~\cite{Amhis:2014hma}\footnote{
The ratio of the effective Fermi constant $G_{\m\t}/G_{e\t}$, $G_{\m\t}/G_{e\m}$ and $G_{e\t}/G_{e\m}$ are
corresponds to $g_\m/g_e$, $g_\t/g_e$ and $g_\t/g_\m$ in Ref.~\cite{Amhis:2014hma}, respectively.}
\begin{align}
 \frac{G_{\mu \tau}}{G_{e \mu}}
=
1.0029 \pm 0.0015
, \quad
 \frac{G_{\mu \tau}}{G_{e \tau}}
=
1.0018 \pm 0.0014
,
\label{eq:constraint_HFAG}
\end{align} 
and their correlation coefficient is 0.48.
We use this bound and Eq.~(\ref{taudecay}) to make constraint on 2HDM\footnote{
At the tree level discussion, we can replace $G_{e\mu}$ and $G_{e\tau}$ with $G_F$,
then we find $G_{\mu \tau}/G_{F} = 1.0023 \pm 0.0012$ or $(G_{\mu \tau}/G_{F})^2 = 1.0046 \pm 0.0025$
by combining Eq.~(\ref{eq:constraint_HFAG}).
However, this treatment does not work for the calculation including one-loop corrections which is given in Sec.~\ref{sec:lu_1loop}.
}.

In Fig.~\ref{fig:taudecay}, we show the $z$ dependence of the ratio of the decay rate given in Eq.~(\ref{taudecay}) (upper two panels) and 
the Michel parameters $\eta$ (lower left) and $\xi$ (lower right).  
First, from the upper panels we can see that the allowed ranges of $z$ are
found to be $z\lesssim 0.003$ and $0.50\lesssim z \lesssim 0.57$. 
Second, from the lower left panel, $z\gtrsim 0.05$ is excluded by the measurement of $\eta$. 
The constraints from the $\xi$ parameter is weaker than that from $\eta$.
Therefore, by combining the first and the second statements, the allowed
region of $z$ is restricted to be $z \lesssim 0.003$
By using $z\simeq 1.88\times 10^{-3}\times (\tan\beta/30)^2\times (300\text{ GeV}/m_{H^\pm}^{})^2$, 
we find that $\tan\beta\gtrsim 38$ is excluded for $m_{H^\pm}=300$ GeV.

\subsubsection{Lepton universality at the one-loop level} \label{sec:lu_1loop}

\begin{figure}[t!]
  \includegraphics[width=0.25\hsize]{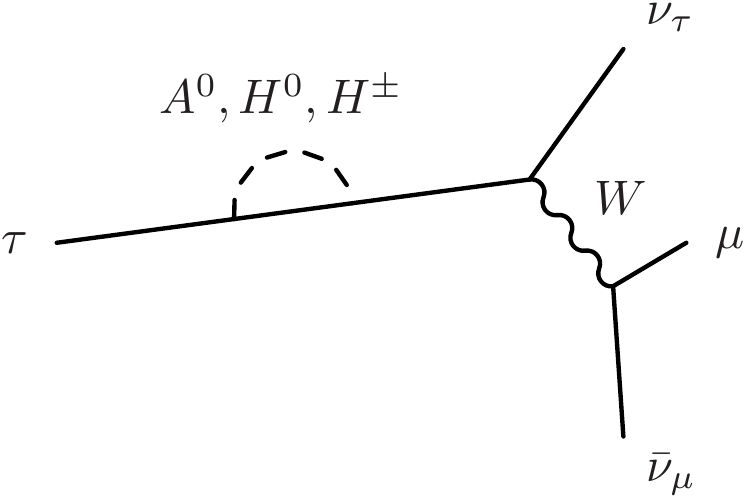}
\qquad
  \includegraphics[width=0.25\hsize]{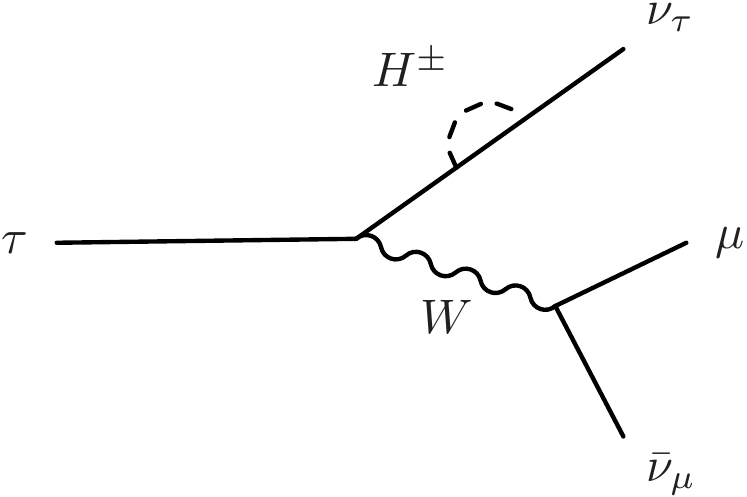}
\qquad
  \includegraphics[width=0.25\hsize]{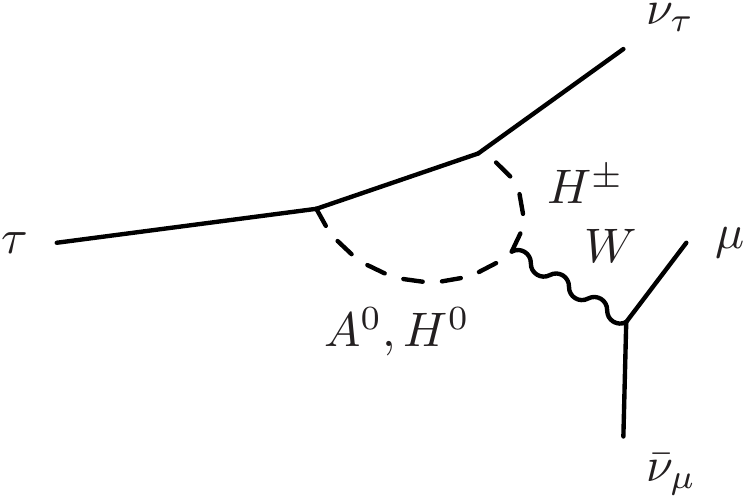}
\caption{
The leading one-loop diagrams for the leptonic tau decay process.
}
\label{fig:loop}
\end{figure}

\begin{figure}[tb]
\centering
  \includegraphics[width=0.35\hsize]{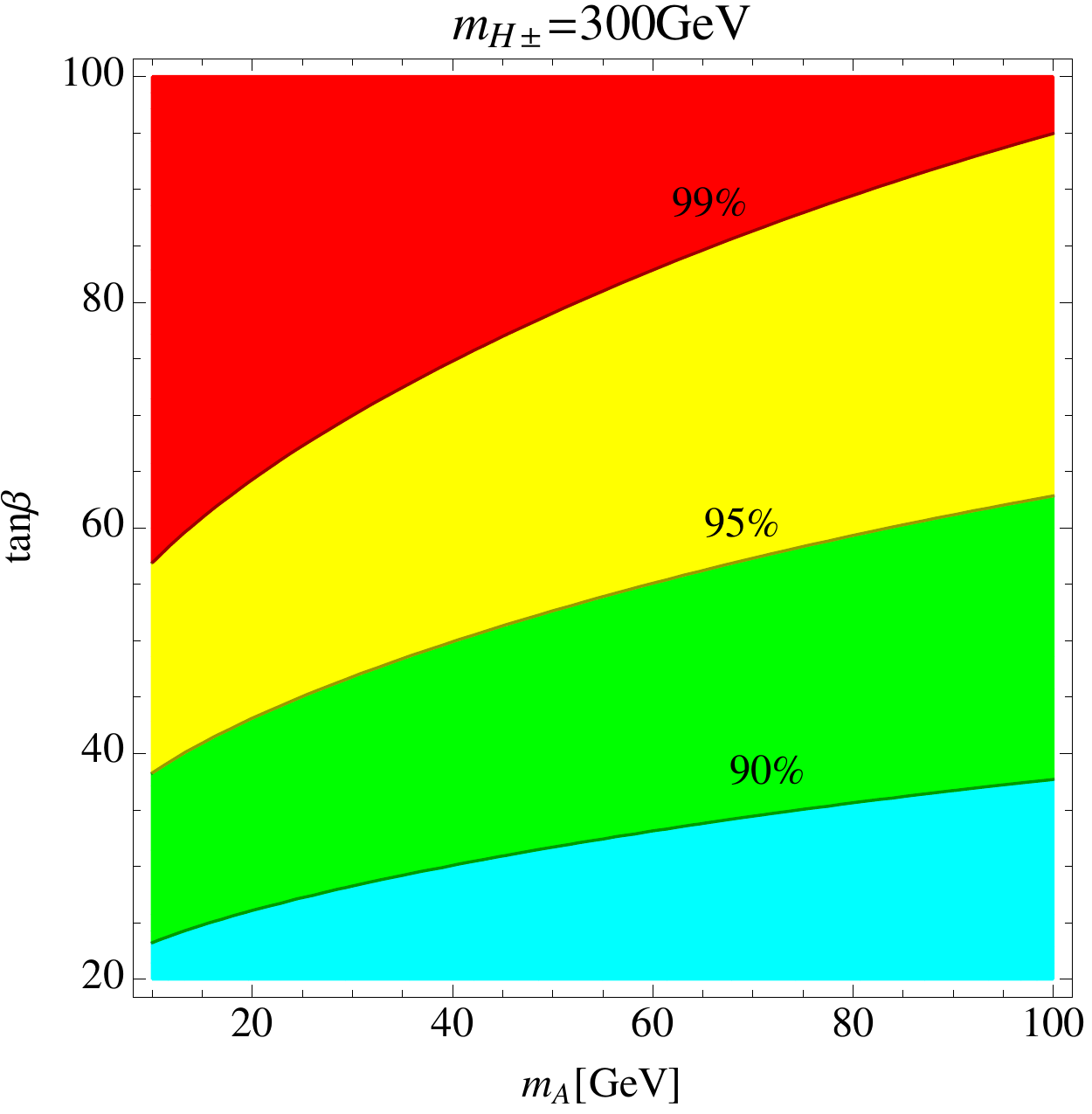}
  \includegraphics[width=0.35\hsize]{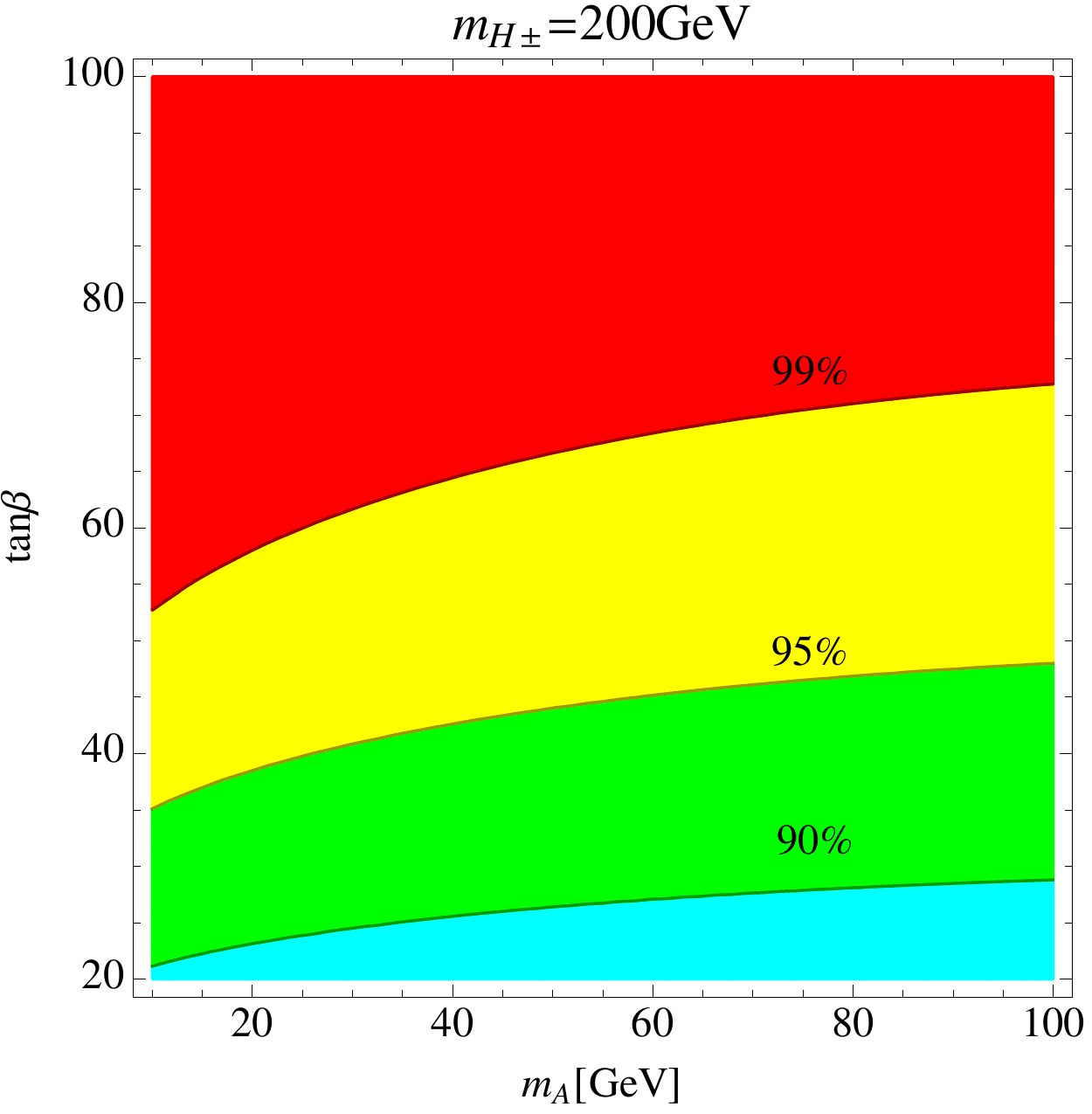}
\\ 
  \includegraphics[width=0.35\hsize]{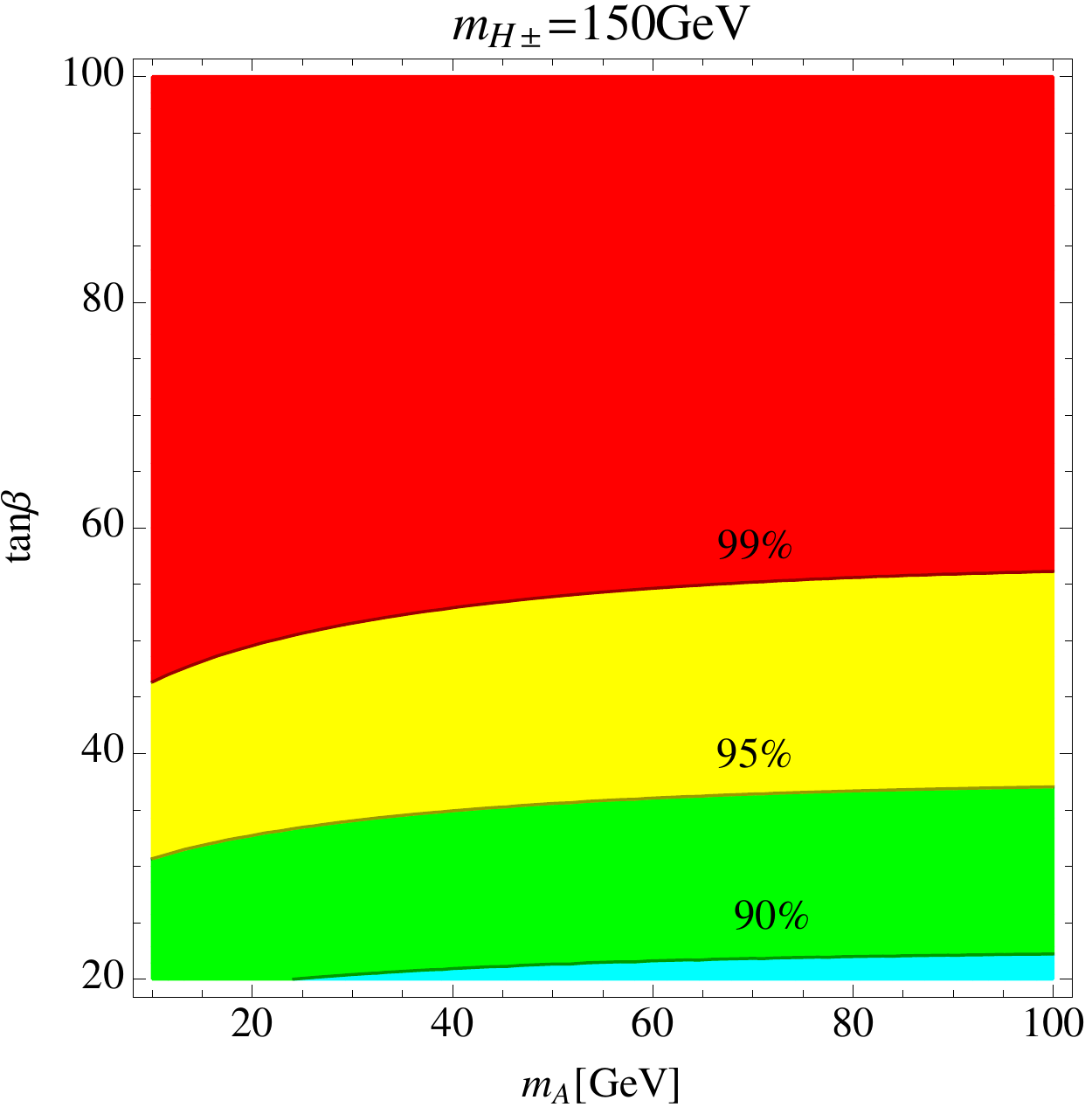}
  \includegraphics[width=0.35\hsize]{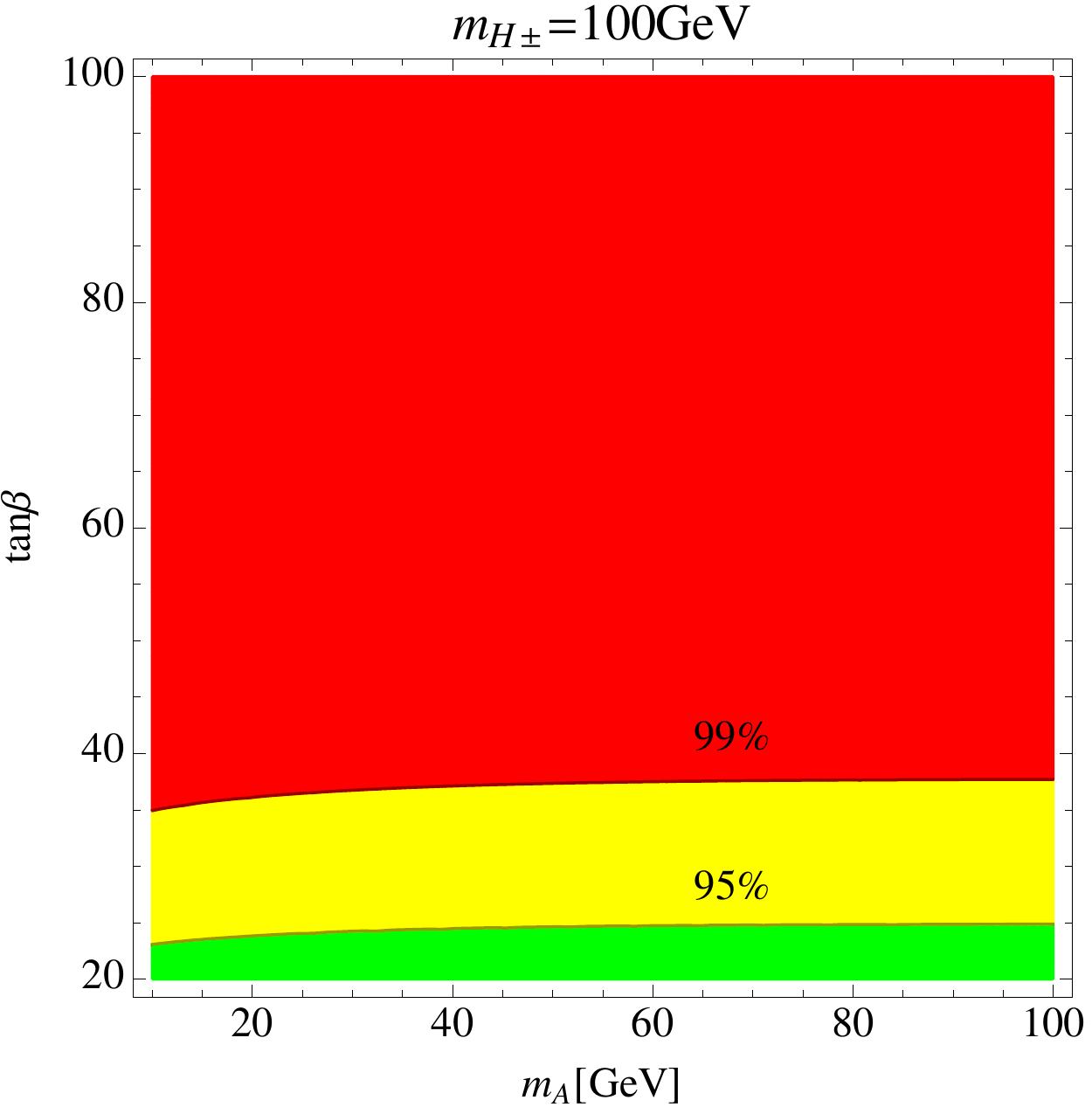}
\caption{
Constraints on the ($m_A^{}, \tan\beta$)-plane by the leptonic tau decay at the one-loop level
for $m_{H^{\pm}}=$300 (upper left), 200 (upper right), 150 (lower left), and 100~GeV (lower right) in the case with  $m_{H^{}} =  m_{H^{\pm}}$. 
The green, yellow, and red regions are excluded at 90\%, 95\%, and 99\%
 C.L., respectively. }
\label{fig:lepton_univ}
\end{figure}

As we discussed in Sec.~\ref{sec:taudecay}, 
the typical size of the $H^\pm$ contribution to the ratio of the tau decay is 
${\cal O}(10^{-2})$ at the tree level as it is seen in Fig.~\ref{fig:taudecay}. 
However, the SM prediction is given at almost the lower edge of the experimental bound, (see Eq.~(\ref{eq:constraint_HFAG})), 
so that the negative contribution to $G_{\m\t}/G_F$ of order $10^{-4}$ is constrained. 
Thus, we focus on the quantum corrections to the process via $W$ exchange diagram.

The dominant contribution arises from the diagrams with picking up two tau Yukawa couplings
which are proportional to $(m_{\tau}^2/v^2) \tan^2\beta$.
We show the diagrams which give the dominant contributions to the process
at the one-loop level in Fig.~\ref{fig:loop}. Other diagrams, such as
box diagrams, are smaller than these contributions and we ignore them in
this analysis. The quantum correction is flavor dependent, and there is
no flavor dependent counter terms in this model, so the correction is
finite. 
We find the contributions from Fig.~\ref{fig:loop} modifies the
$W$-$\tau$-$\nu_{\tau}$ couplings
\begin{align}
 g_{W \tau \nu} 
\to
 g_{W \tau \nu} 
\left(
1 + \delta g
\right)
,
\end{align} 
where
\begin{align}
\delta g
=&
\frac{1}{(4\pi)^2}
\frac{m_{\tau}^2}{v^2}
\tan^2 \beta
\left[
1
+
\frac{m_{H^{\pm}}^2 + m_{A}^2}{4 (m_{H^{\pm}}^2 - m_{A}^2)}
\ln \frac{m_{A}^2}{m_{H^{\pm}}^2}
+
\frac{m_{H^{\pm}}^2 + m_{H}^2}{4 (m_{H^{\pm}}^2 - m_{H}^2)}
\ln \frac{m_{H}^2}{m_{H^{\pm}}^2}
\right]
.
\end{align}
In $m_{A} \ll m_{H^{\pm}} \sim m_{H^{0}}$ case
\begin{align}
\delta g
=&
\frac{1}{(4\pi)^2}
\frac{m_{\tau}^2}{v^2}
\tan^2 \beta
\left[
\frac{1}{2}
+
\frac{1}{4}
\ln \frac{m_{A}^2}{m_{H^{\pm}}^2}
+
{\cal O}
\left(
\frac{m_{A}^2}{m_{H^{\pm}}^2}
\ln \frac{m_{A}^2}{m_{H^{\pm}}^2}
\right)
\right]
.
\end{align}
Finally, we find
\begin{align}
 \Gamma
=&
 \frac{1}{96 \pi^3}
m_{\tau} \omega^4
 G_F^2
\Biggl[
\sqrt{1 - x_0^2}
\Bigl(
  2 (4 (1 + \delta g)^2 + z^2)
- 8 x_0 z (1 + \delta g)
\nonumber\\
& \qquad \qquad \qquad \qquad \qquad \qquad
- 5 x_0^2 ( 4 (1 + \delta g)^2 + z^2)
- 16 x_0^3 z (1 + \delta g)
\Bigr)
\nonumber\\
&
\qquad \qquad \qquad \quad 
+
3 x_0^3
\left(
8 z (1 + \delta g) + x_0 (4 (1 + \delta g)^2 + z^2)
\right)
\ln \frac{1 + \sqrt{1 - x_0^2}}{x_0}
\Biggr]
,
\label{eq:1loop_lepton}
\end{align}
where $\omega$ and $x_0$ are defined in Sec.~\ref{sec:taudecay}.
%
Using Eqs.~(\ref{taudecay}) and (\ref{eq:1loop_lepton}), 
we find that the large $\tan\beta$ region and light charged Higgs region
are disfavored, see Fig.~\ref{fig:lepton_univ}. This constraint is the most severe for the explanation
of the muon $g-2$ anomaly, as we will see in Sec.~\ref{sec:muon_g-2}

\subsection{Triviality bound}
 
In order to avoid the constraints from the various observables, 
we need to take large mass differences between $A$ and $H^{\pm}$, and
$A$ and $H$. As a result, 
the Higgs quartic couplings are as large as ${\cal O}(1)$.  
Such a large coupling can be grown up in a certain energy scale, 
and it becomes too strong to rely on the perturbative calculation. 
We thus take into account the triviality bound in which we require that 
all the Higgs quartic couplings do not exceed a certain value until a given energy scale. 

We calculate the $\beta$-functions up to the two loop level for the RGE by using \texttt{SARAH}~\cite{Staub:2013tta}, and run the couplings to higher energies. 
We treat the coupling values at the tree level as the input
parameters at $\mu = m_t$.
We define $\lambda_{\text{max}} = \text{max}\{|\lambda_1|, |\lambda_2|,
|\lambda_3|, |\lambda_4|, |\lambda_5| \}$ and the cutoff scale of the model, $\Lambda$, as
$\lambda_{\text{max}}(\Lambda) = 4\pi$ or $\sqrt{4 \pi}$.

\begin{figure}[tb]
  \includegraphics[width=0.50\hsize]{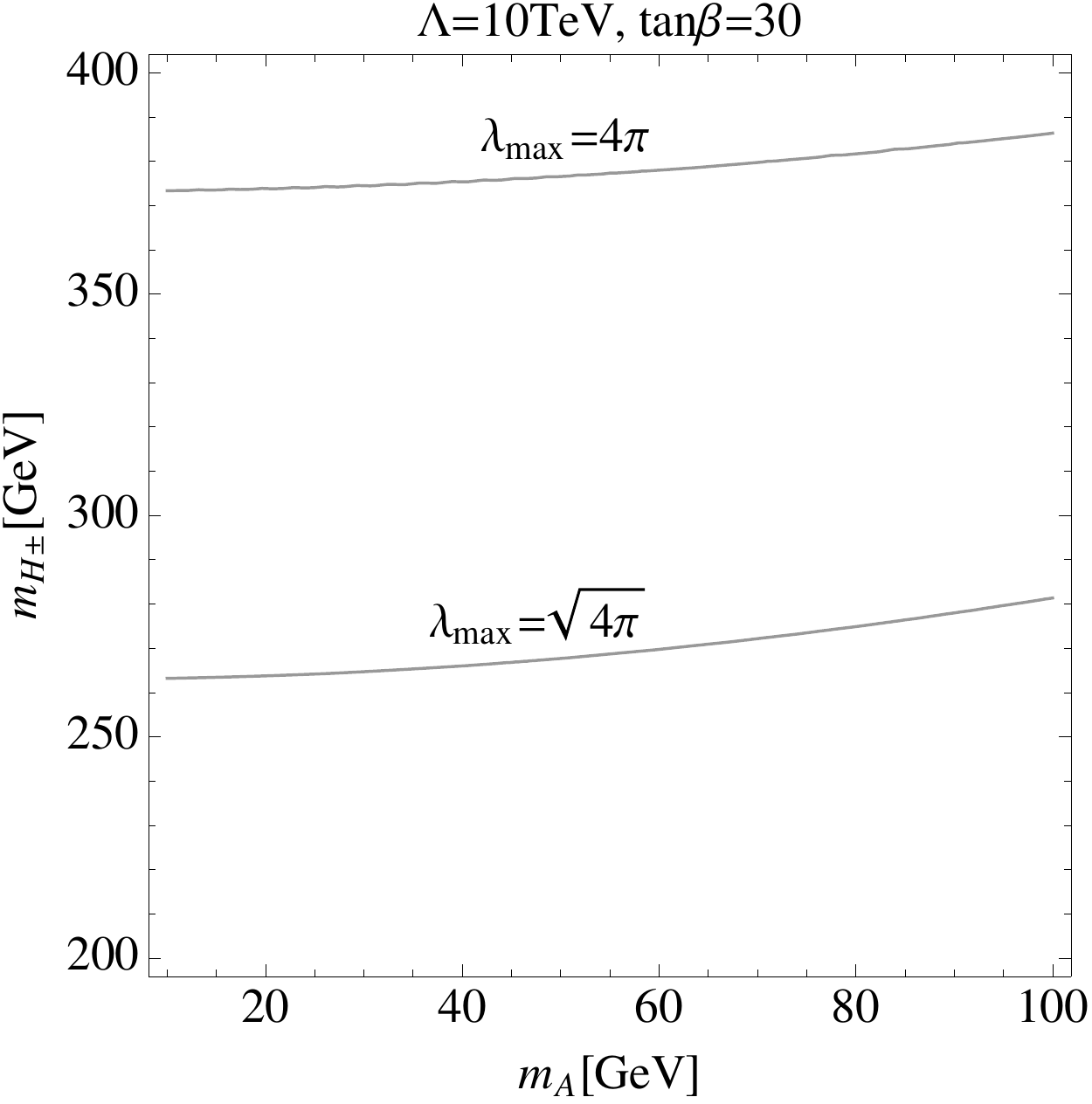}
  \includegraphics[width=0.50\hsize]{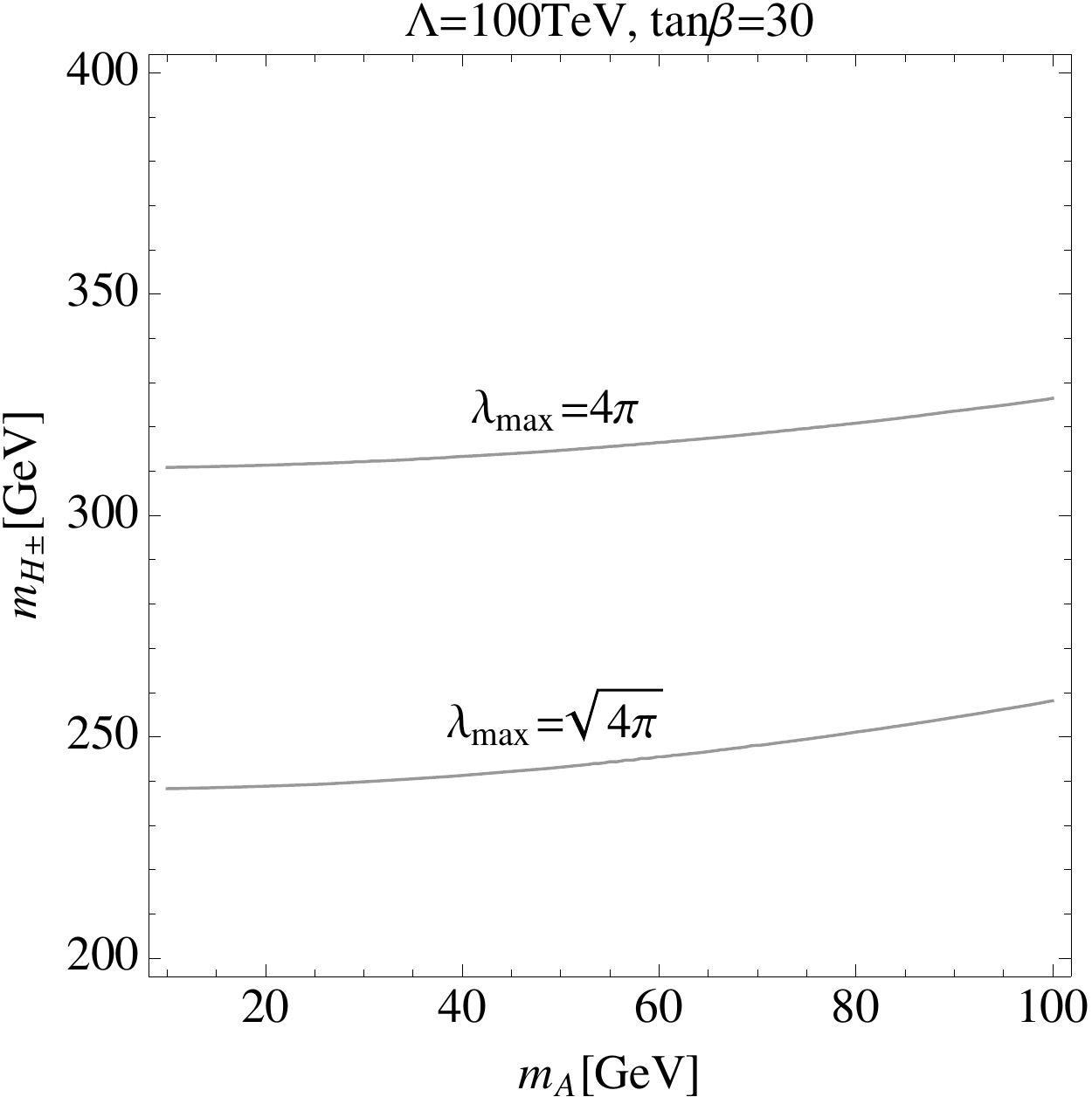}
\caption{
The triviality bound with $\Lambda =$ 10~TeV (left) and $\Lambda =$
 100~TeV (right). The numbers shown in the panel are 
 $\lambda_{\text{max}} = 4\pi $ and $\sqrt{4 \pi}$.
}
\label{fig:triviality}
\end{figure}

The result for $\Lambda = 10$~TeV with requiring $\lambda_{\text{max}}
< 4\pi$ is shown in the left panel in Fig.~\ref{fig:triviality}.
We find that $m_{H^{\pm}} \lesssim 370$~GeV for $m_A \simeq 20$~GeV is
required for $\Lambda ~ 10$~TeV. 
This constraint on $m_{H^\pm}^{}$ is stronger than the one from the perturbative unitarity bound using Eqs.~(\ref{pv1})-(\ref{pv2}), \textit{i.e.}, 
$m_{H^{\pm}} \lesssim 700$~GeV. We check that this result is consistent with that given in~\cite{Gorczyca:2011rs}. 
If we require $\lambda_{\text{max}} < \sqrt{4\pi}$ instead of 
$\lambda_{\text{max}} < 4\pi$, 
then the bound becomes stronger.
In such a case, we find $m_{H^{\pm}} \simeq 260$~GeV is required. 
Here we take $\tan \beta =30$ at $\mu = m_t$ as the input value, but the result is
insensitive for $\tan \beta$. 
We also plot the case for $\Lambda = 100$~TeV in the right panel in
Fig.~\ref{fig:triviality}. The bound is stronger than $\Lambda = 10$~TeV
case. We find
$m_{H^{\pm}} \lesssim 310$~GeV  (240~GeV) for
$\lambda_{\text{max}}=4\pi \ (\sqrt{4\pi})$.
Our parameter choice here is the same as in Sec.~\ref{sec:muon_g-2}.

\section{Muon $g-2$}\label{sec:muon_g-2}

\begin{figure}[p]
\centering
\includegraphics[width=0.48\hsize]{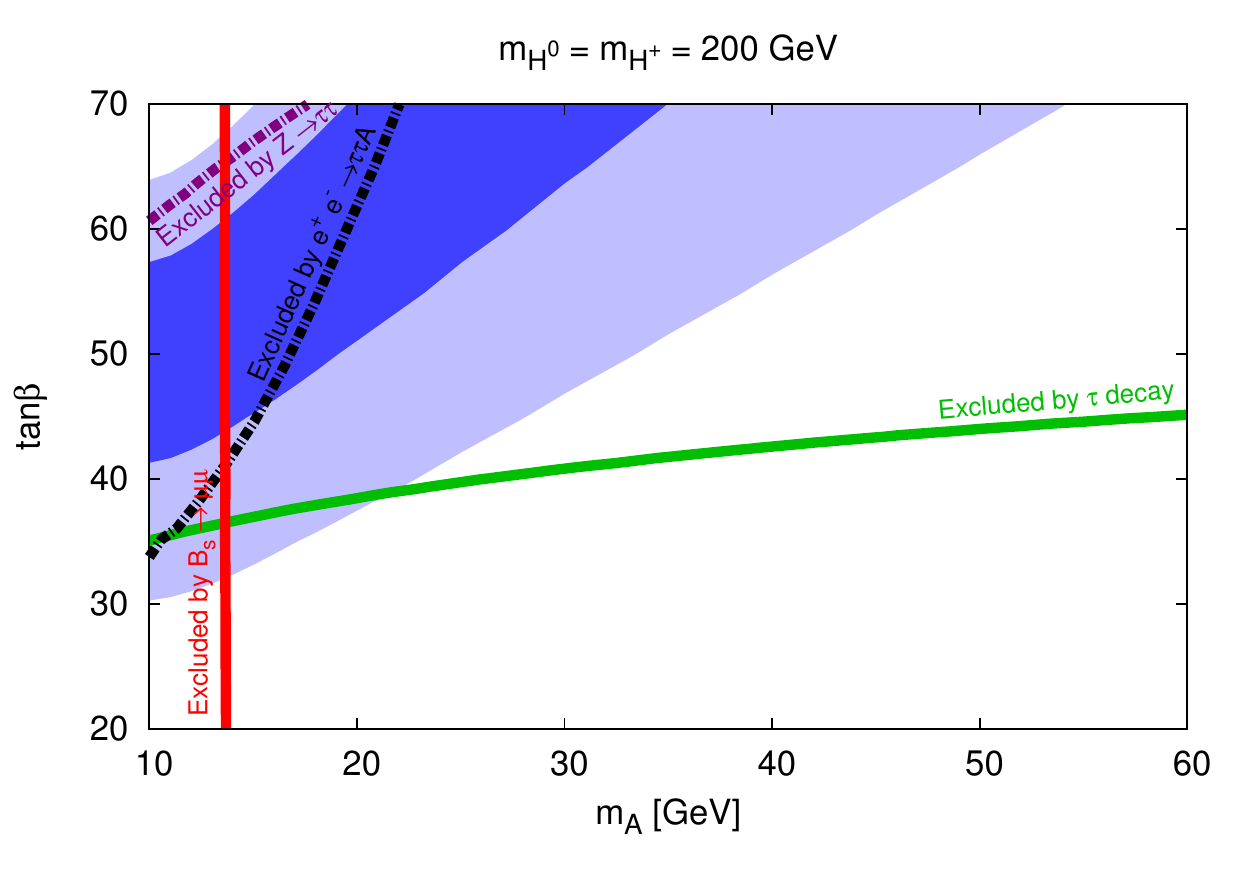} 
\includegraphics[width=0.48\hsize]{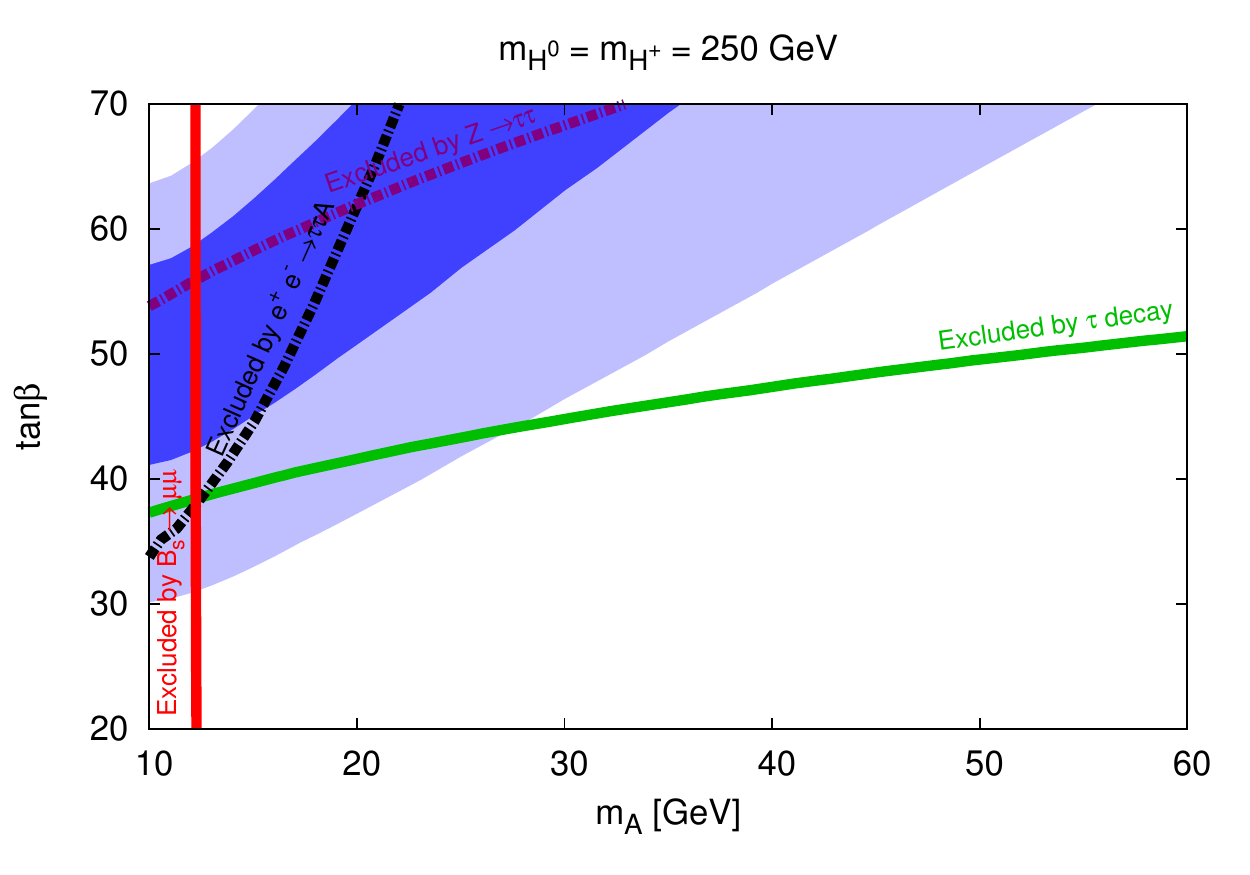} \\
\includegraphics[width=0.48\hsize]{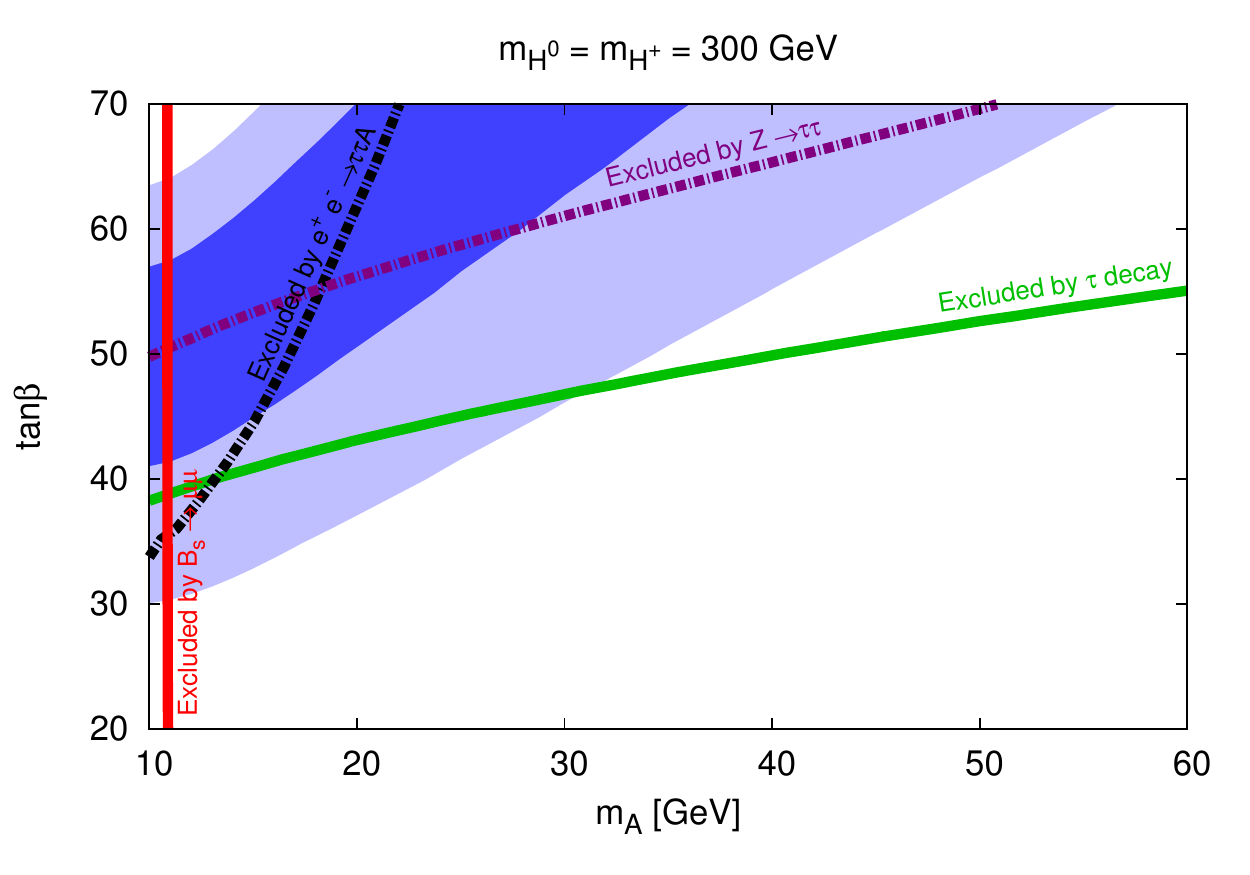} 
\includegraphics[width=0.48\hsize]{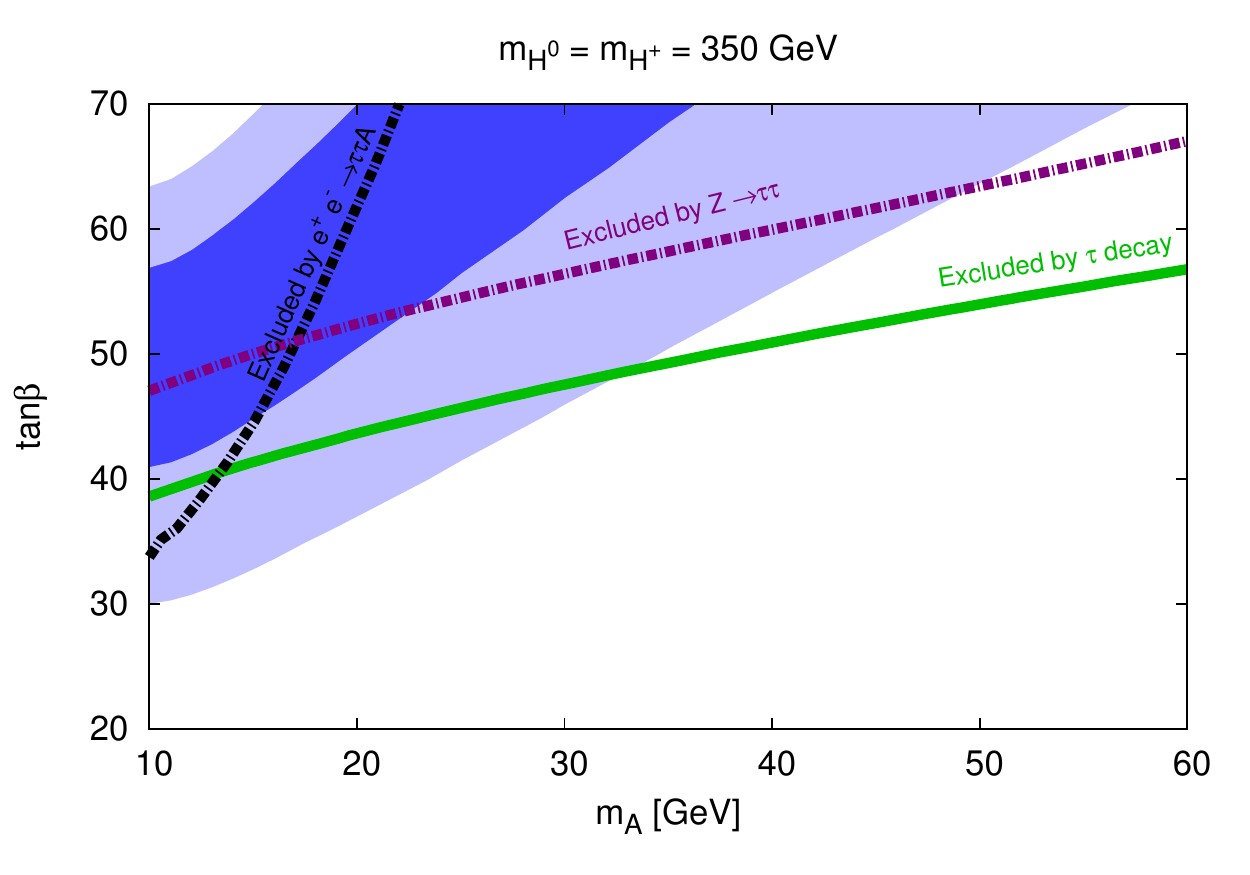} 
\caption{Results in the Type-X 2HDM in the case of $\l_{hAA}=0$ and $\l_1 = 0.1$.
Dark and light blue shaded regions can explain the muon $g-2$ anomaly \cite{Hagiwara:2011af} at the $1\s$ and $2\s$ levels, respectively.
We take $m_{H^\pm}(=m_H^{})=$200, 250, 300 and 350 GeV in the upper-left, upper-right, lower-left and lower-right panels, respectively.
The left region from the red line is excluded by the measurement of $B_s \to \m\m$.
The above regions of green, black, purple line are excluded by the $\t$ decay, the direct search at the LEP and the $Z\to\t\t$ decay, respectively.
All of the exclusions are given at the 95\% C.L.
}\label{fig:typeX}
\end{figure}

We show the numerical results for the muon $g-2$ under all the constraints discussed in the previous section. 
We calculate the muon $g-2$ by using {\tt 2HDMC 1.6.4} \cite{Eriksson:2009ws} which contains 
the one-loop diagrams \cite{Dedes:2001nx} and the two-loop Barr-Zee
diagrams \cite{Chang:2000ii} as shown in Fig.~\ref{fig:amu}. 
In the Type-X
2HDM, the contributions from the one-loop diagrams and the Barr-Zee
diagrams are comparable and have opposite sign. Thus, we have to take
into account both contributions. 
The contributions from the beyond the SM part should be 
sizable to solve the muon $g-2$ anomaly,
and thus at least one of the new particles has to be light. The lower mass
bound on $H^\pm$ is order of 100~GeV, so that it can not be
arbitrary small. 
The effect from $H$ gives a destructive contribution to the SM one, so that it makes 
the situation becomes even worse if $H$ is light. 
On the other hand,
the effect from $A$ gives the constructive contribution, and it 
makes the discrepancy small. Therefore, $A$ is required to be lighter than $H$ and
$H^{\pm}$ in order to solve the muon $g-2$ anomaly.
%


The Higgs sector in the 2HDM has eight parameters.
Two of them are fixed to reproduce the SM parameters, \textit{i.e.}, $G_F=2^{-1/2}v^{-2}=1.166379\times 10^{-5}~{\rm GeV}^{-2}$ and $m_h=125~{\rm GeV}$.
To suppress ${\rm Br}(h\to AA)$, we set $\l_{hAA}=0$.
In addition, we take $m_H^{}=m_{H^\pm}$ to avoid a large contribution to $\Delta T$. 
Furthermore, we fix $\l_1=0.1$. This $\lambda_1$ value is realized by
taking $M^2\simeq m_H^2$ in the large $\tan\beta$ case, and the fixing
the value of $\lambda_1$ is not significant to the result in the following.
Therefore, we have three remaining parameters which can be expressed as $\tan\b$, $m_A$ and  $m_{H^\pm}$. 
We note that in this parametrization, $\sin(\beta-\alpha)$ is given as the output parameter, which  
is determined via Eq.~(\ref{lam_haa}). 
As it is seen in Eq.~(\ref{sin_app}), 
$1-\sin(\beta-\alpha)$ is suppressed by $1/\tan^2\beta$ in the large $\tan\b$ case, so that 
$h$ behaves as the SM-like Higgs boson.

In Fig.~\ref{fig:typeX}, we show the prediction of $a_\mu$ on the $m_A^{}$-$\tan\beta$ plane in each fixed value of $m_{H^\pm}^{}$, \textit{e.g.}, 
$m_{H^\pm}^{}=200$ (upper-left), 250 (upper-right), 300 (lower-left) and 350 GeV (lower right). 
In the dark and light shaded regions, $a_\mu$ can be explained at the 1$\sigma$ and 2$\sigma$ levels, respectively. 
We find that the measurement of $\t\to\m\n_\m\n_\t$ gives the stringent constraint on the parameter space in the $m_A^{}$-$\tan\beta$ plane. 
This constraint is getting stronger in the cases of smaller $m_{H^\pm}$ or $m_A^{}$ and larger $\tan\beta$. 
For example, in the case of $m_{H^\pm}=200$ GeV, 
$\tan\beta \gtrsim 35$ is excluded at the 95\% C.L.~in the case of $m_A=20~{\rm GeV}$.
For $m_A^{}\simeq 10$ GeV, ${\rm Br}(B_s \to \m^+\m^-)$ gives additional excluded regions which are not excluded by the measurement of the tau decay. 
The $\tan\beta$ dependence in ${\rm Br}(B_s \to \m^+\m^-)$ is negligible as we expected in the discussion in Sec.~\ref{sec:bsmm}. 
In most of the parameter region shown in this figure, 
the deviation of ${\rm Br}(B_s \to \m^+\m^-)$ from the SM value is ${\cal O}(1)$~\%. 
The constrained from $Z\to \tau\tau$ is weaker than that from the tau decay in all the parameter regions shown in this figure. 

Consequently, the parameter region which can explain the $g-2$ anomaly at the 1$\sigma$ level is excluded by 
the measurement of the tau decay at the 95\% C.L., 
and at best we can explain the anomaly at the 2$\sigma$ level. 
The typical parameters to explain muon $g-2$ anomaly at the 2$\sigma$ level is
$10 \lsim m_A \lsim 30~{\rm GeV}$, $200 \lsim m_{H^\pm}^{} \lsim 350~{\rm GeV}$ and $30 \lsim \tan\b \lesssim 50$.
In this parameter space, however, the region with $m_{H^\pm} \lesssim 270$ GeV
has tension with the signal strength for the $h\to \tau\tau$ mode 
measured at the LHC as we will see in the next section.

\section{Impact on the Higgs phenomenology at collider experiments}\label{sec:phenomenology}

In the previous section, we have seen that the relatively light extra Higgs bosons and large $\tan\b$ are 
favored to explain the  muon $g-2$ anomaly.
Such a light particle can be directly discovered at the LHC Run-II and the International Linear Collider (ILC).
Furthermore, the precise measurement of the property of the SM-like Higgs boson $h$ will 
give an indirect probe of this scenario.
In this section, we first discuss the decay and production of the extra Higgs bosons at the LHC, 
and then we investigate how the property of the SM-like Higgs boson is modified 
in the favored parameter set indicated by the muon $g-2$ in the Type-X 2HDM.

Throughout this section, we consider the case of 
\begin{align}
&200~\text{GeV} \leq m_{H^\pm}\leq 400~\text{GeV},~~m_H^{}=M=m_{H^\pm},~~
10~\text{GeV} \leq m_{A}^{}\leq 30~\text{GeV}, \notag\\
&30 \leq \tan\beta \leq 50,~~ \tan(\beta-\alpha) = \frac{M^2-m_h^2}{2M^2-2m_A^2-m_h^2}(\tan\beta-\cot\beta). \label{scenario}
\end{align}
In addition, the following values of the SM parameters are used~\cite{PDG2014,Koide}:
\begin{align}
&m_Z=91.1876~\text{GeV},~
m_W=80.385~\text{GeV},~
G_F=1.1663787\times 10^{-5}~\text{GeV}^{-2}, \notag\\
&m_t=173.07~\text{GeV},~          
m_b = 3.0~\text{GeV},~
m_c = 0.677~\text{GeV},~
m_\tau=1.77684~\text{GeV}. 
\end{align}
$m_c$ and $m_b$ are evaluated at the $m_Z$ scale in $\overline{\rm MS}$ scheme, which are taken from Ref.~\cite{Koide}. 
Other parameters are taken from Ref.~\cite{PDG2014}.
The mass observed Higgs boson is taken to be 125 GeV. 

\subsection{Phenomenology of the extra Higgs bosons}\label{sec:extra_Higgs}

First, we discuss the branching fraction of the extra Higgs bosons in the parameter set given in Eq.~(\ref{scenario}).
For the CP-odd Higgs boson $A$, 
only the $A\to \tau\tau$ and $A\to b\bar{b}$ modes are allowed at the tree level. 
Since the decay rate of the former (latter) channel is enhanced (suppressed) by $\tan^2\beta$ ($\cot^2\beta$) in the Type-X 2HDM, 
the branching fraction of $A\to \tau\tau$ becomes almost 100\% in our scenario. 
Similar enhancement happens in the decay rates of $H\to \tau\tau$ and $H^\pm \to \tau^\pm\nu$. 
At the same time, the other modes $H\to AZ$ and $H^\pm \to AW^\pm$ are also important 
because of the large mass difference between $H/H^\pm$ and $A$. 
Therefore, the following decay modes should be taken into account:
\begin{align}
&H^\pm \to \tau^\pm\nu,\quad H^\pm \to AW^\pm, \quad
H \to \tau\tau,\quad H \to AZ,\quad A \to \tau\tau. 
\end{align} 
The formulae for the decay rates are given in Appendix~A. 
We here note that the $H\to AA$ and $H\to hh$ decays also open whose decay rates are determined by the trilinear $HAA$ and $Hhh$ couplings, respectively. 
However, these couplings are proportional to $\cos(\beta-\alpha)$, and its magnitude is suppressed by $\cot\beta$ in the large $\tan\beta$ case. 
Thus, these decay modes are not important in our scenario. 

\begin{figure}[t]
\centering
\includegraphics[width=0.48\hsize]{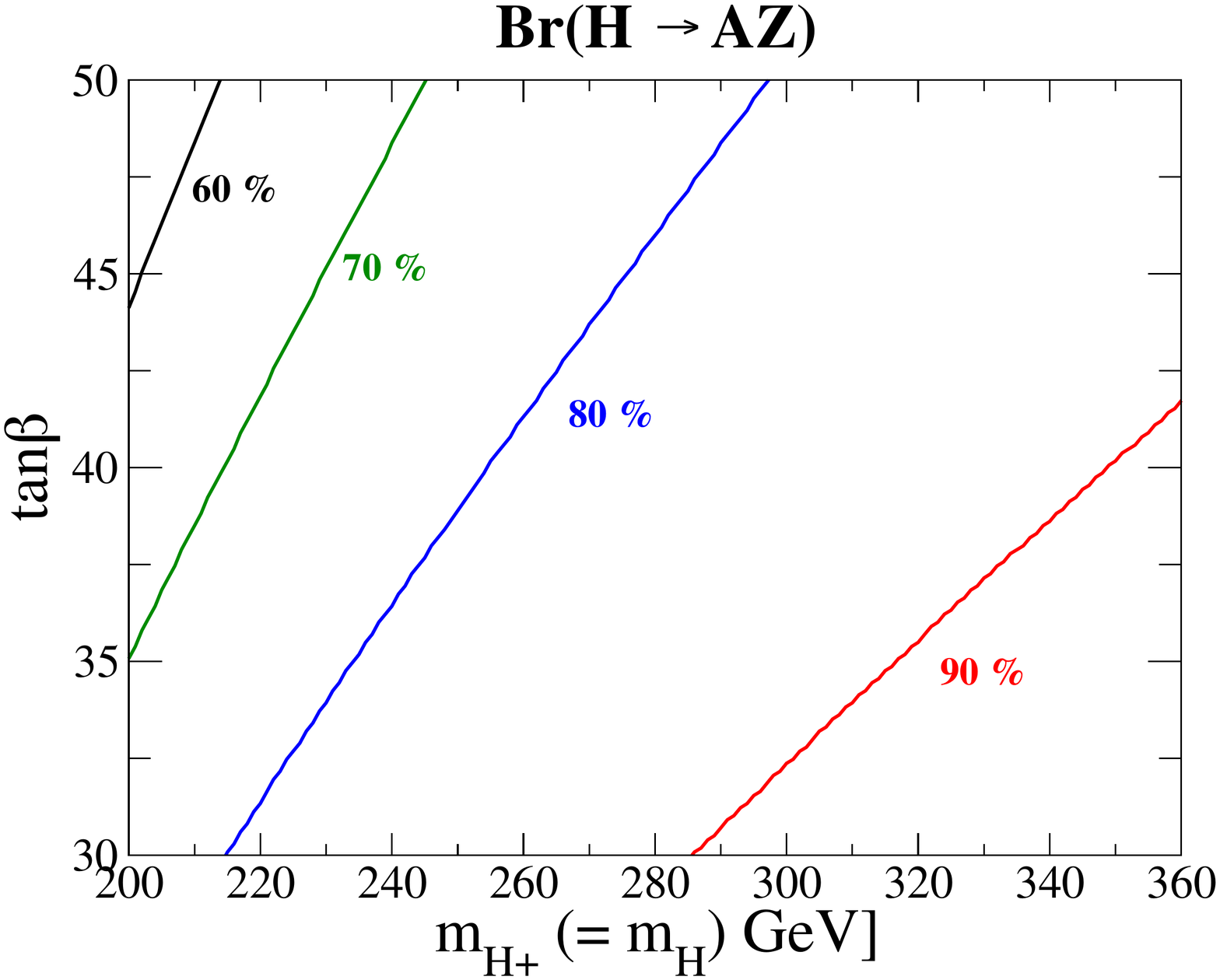} 
\includegraphics[width=0.48\hsize]{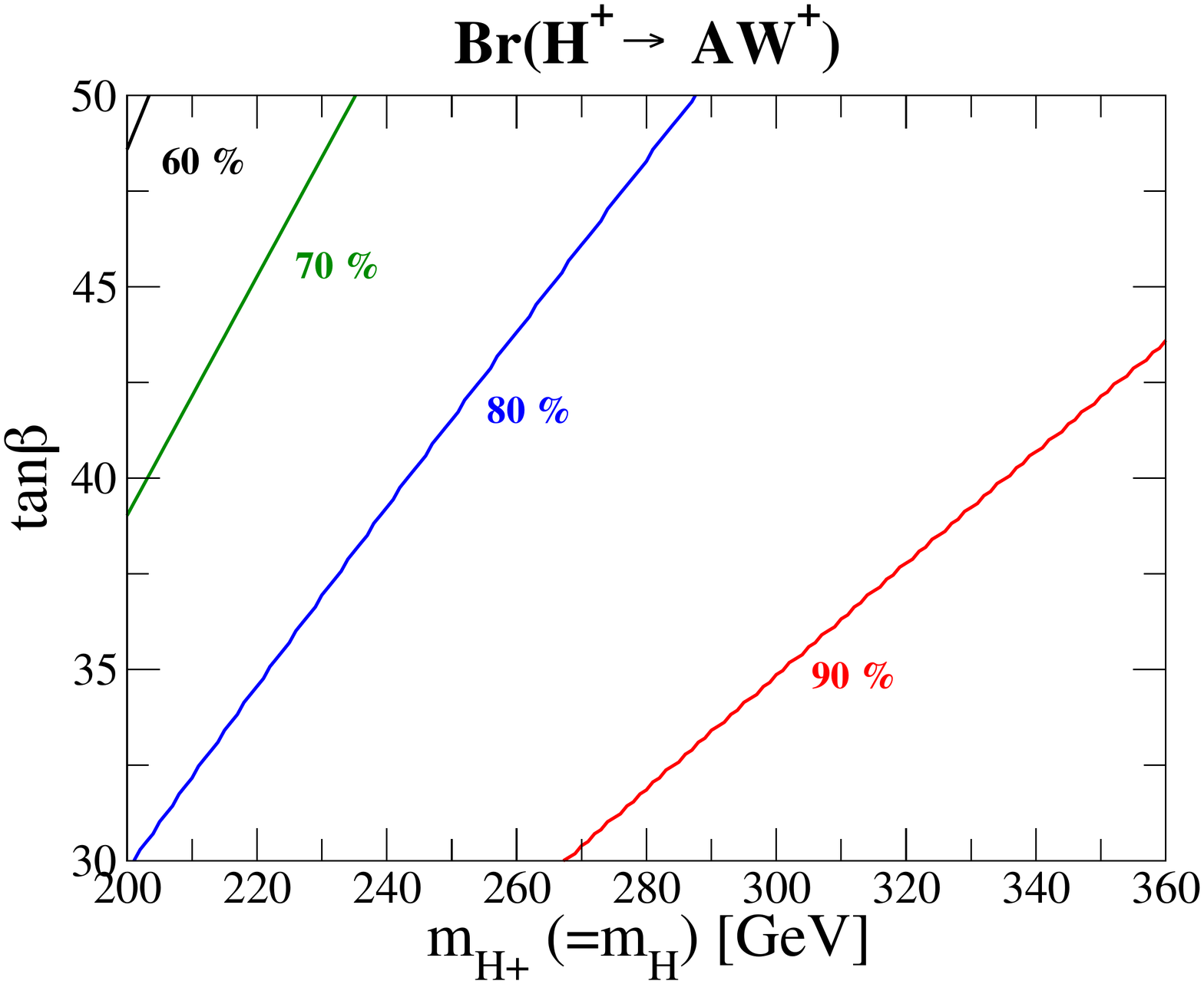} 
\caption{Contour plots for the branching fractions of $H\to AZ$ (left panel) and $H^+ \to AW^+$ (right panel)
on the $m_{H^\pm}$-$\tan\beta$ plane. 
We take $m_H^{}=m_{H^\pm}$ and $m_A^{}=20$ GeV. 
Br($H\to AZ)+$Br($H\to \tau\tau)\simeq 1$ and 
Br($H^+\to AW^+)+$Br($H^+\to \tau^+\nu)\simeq 1$ are satisfied.  }\label{fig:decay_H}
\end{figure}

In Fig.~\ref{fig:decay_H}, we show the contour plots for the branching fractions of $H\to AZ$ (left panel)
and $H^+ \to AW^+$ on the $m_{H^\pm}$-$\tan\beta$ plane. 
The $H\to \tau\tau$ and $H^+\to \tau^+\nu$ modes fulfill almost the remaining branching ratios for $H$ and $H^\pm$, respectively. 
In the large $m_{H^\pm}$ and small $\tan\beta$ region, 
the branching fractions of $H\to AZ$ and $H^+ \to AW^+$ are getting larger. 
For example, we obtain Br($H\to AZ)>70~(90)$\%
in the case of $\tan\beta \lesssim 35$ with $m_{H}^{}=200$ GeV ($\tan\beta \lesssim 33$ with $m_{H}^{}=300$ GeV), and 
Br($H^+\to AW^+)>70~(90)$\%
in the case of $\tan\beta \lesssim 38$ with $m_{H}^{}=200$ GeV ($\tan\beta \lesssim 35$ with $m_{H^\pm}^{}=300$ GeV). 

Next, we discuss the production process of the extra Higgs bosons at the LHC. 
As we mentioned in Sec.~\ref{sec:LHC1}, the production processes via the Yukawa interaction cannot be used in the large $\tan\beta$ case in the Type-X 2HDM. 
Therefore, the following electroweak processes give the dominant production mode:
\begin{align}
&pp \to Z^*/\gamma^* \to H^+H^-,\quad 
pp \to Z^* \to HA,\notag\\ 
&pp \to W^* \to H^\pm H,\quad 
pp \to W^* \to H^\pm A.  \label{ew}
\end{align}
The analytic expressions for the parton level cross section are given in Appendix~B. 
We find that the cross sections are determined by the masses of the extra Higgs bosons and $\sin(\beta-\alpha)$, 
and they do not depend on the type of Yukawa interactions.  
By using {\tt CalcHEP}~\cite{calchep} with {\tt CTEQ6L}~\cite{cteq} parton distribution functions, 
the cross sections are calculated in Table~\ref{tab_xs}.  
In this calculation, we neglect the small deviation in $\sin(\beta-\alpha)$ from unity. 
Because of the small $m_A^{}$, the cross sections of $pp \to H^\pm A$ and $pp\to HA$ are 
relatively large as compared to those of $pp \to H^+ H^-$ and $pp\to H^\pm H$. 
We note that the cross section for $H^+H/H^+A$ is about twice larger than that for $H^-H/H^-A$, because 
the parton luminosity of $u\bar{d}$ in the initial proton is larger than that of $\bar{u}d$. 

Combining the discussions of the decay and the production of the extra Higgs bosons, 
we can consider the following processes: 
\begin{align}
\begin{array}{ll}
pp \to H^\pm A/H^\pm H \to \tau^\pm \nu \tau^+\tau^-, &pp \to H^\pm A/H^\pm H \to 4\tau+W^\pm, \\
pp \to HA \to 4\tau,  &  pp \to H A  \to 4\tau+Z. 
\end{array}
\end{align}
The $H^+H^-$ production may not be so useful 
for the feasibility study of the extra Higgs bosons as compared to the above processes because of the small cross section as seen in Table~\ref{tab_xs}. 
%
The cross sections of 4$\tau$ ($\sigma_{4\tau}$), $3\tau$~($\sigma_{3\tau}$), $4\tau$ plus $W$~($\sigma_{4\tau W}$) and
$4\tau$ plus $Z$~($\sigma_{2\tau Z}$) can be estimated as follows:
\begin{align}
\sigma_{4\tau} &\simeq \sigma_{HA}\times \text{Br}(H\to \tau\tau),  \label{sig1}\\
\sigma_{3\tau} &\simeq (\sigma_{H^+A}+\sigma_{H^-A})\times \text{Br}(H^\pm \to \tau^\pm\nu) \notag\\
&+ (\sigma_{H^+H}+\sigma_{H^-H})\times \text{Br}(H^\pm \to \tau^\pm\nu)\times \text{Br}(H \to \tau\tau),  \label{sig2}\\
\sigma_{4\tau W} &\simeq (\sigma_{H^+A}+\sigma_{H^-A})\times \text{Br}(H^\pm \to AW^\pm) \notag\\
&+ (\sigma_{H^+H}+\sigma_{H^-H})\times \text{Br}(H^\pm \to AW^\pm )\times \text{Br}(H \to \tau\tau),  \label{sig3}\\
\sigma_{4\tau Z} &\simeq \sigma_{HA}\times \text{Br}(H\to AZ), \label{sig4} 
\end{align}
where we used Br$(A\to \tau\tau)\simeq 100\%$. 
The cross sections of the above processes are also shown in Table~\ref{tab_xs} in the case of $\tan\beta=35$.  

The signal and background for the four and three tau final states (without a gauge boson) were studied in Ref.~\cite{Kanemura:2011kx} in the Type-X 2HDM. 
It was clarified that the main background for these processes can be significantly reduced by
requiring the high multiplicity of charged leptons and tau-jets with appropriate kinematical cuts in the final state.
In this paper, we only show the signal cross sections of the above mentioned processes as given in Table~\ref{tab_xs}.
Although the detailed background simulation must be necessary to clarify the feasibility to detect the signal events, 
such a study is  beyond the scope of this paper.

\begin{table}
\centering
\begin{tabular}{|c||cccccc||cccc|}
\hline
$m_{H^\pm}$ [GeV] & $\sigma_{H^+H^-}$ & $\sigma_{H^+H}$ &$\sigma_{H^-H}$ & $\sigma_{H^+A}$ & $\sigma_{H^-A}$ & $\sigma_{AH}$ 
& $\sigma_{4\tau}$ &$\sigma_{3\tau}$ &$\sigma_{4\tau W}$ &$\sigma_{4\tau Z}$ \\\hline\hline
200  & 18.6 &22.0&11.3&116 &67.0&101& 29.3 &50.1&143 &70.7\\\hline
250  & 8.0  &9.7 &4.7 &53.5&29.5&45.1&7.2 &12.8&72.5& 37.4\\\hline
300  & 3.9  &4.8 &2.3 &28.2&14.9&23.2&2.3 &4.3 &39.4& 20.6\\\hline
350  & 2.1  &2.6 &1.1 &16.2&8.2 &13.0&0.9  &1.7 &22.9& 12.0\\\hline
\end{tabular}
\caption{Cross sections of the electroweak production processes expressed in Eq.~(\ref{ew}), 
and those of the multi-tau processes expressed in Eqs.~(\ref{sig1})-(\ref{sig4})
at $\sqrt{s}=14$ TeV in the unit of fb. 
We take $m_A^{}=20$ GeV, $m_H^{}=m_{H^\pm}^{}$, $\sin(\beta-\alpha)=1$ and $\tan\beta=35$. 
}
\label{tab_xs}
\end{table}

\subsection{Phenomenology of the SM-like Higgs boson}

Another important impact on the Higgs phenomenology in our scenario is found in the property of the SM-like Higgs boson $h$. 
Because the properties of $h$; \textit{e.g.},  
the width, the branching fractions, and the couplings will be precisely measured at future collider experiments such as 
the LHC Run-II, the high luminosity LHC, and the ILC~\cite{snowmass}, it must be quite important to study 
the deviation in the property of $h$ from the SM prediction. 
In particular, studying the pattern of the deviation in the various $h$ couplings can be a powerful tool to determine
the structure of the Higgs sector\footnote{In Ref.~\cite{fingerprint}, 
the pattern of the deviation was investigated in various extended Higgs sectors; \textit{e.g.}, 
models with isospin singlets, doublets and triplets at the tree level. 
For example, it was shown that the four types of Yukawa interactions in the 2HDM can be well discriminated by measuring 
the correlation between the deviation in $hd\bar{d}$ and $h\ell\ell$ couplings~\cite{fingerprint}. 
In addition, it was clarified in Ref.~\cite{yukawa_loop} that 
even if we take into account the one-loop corrections to the $hf\bar{f}$ couplings, 
discrimination of the 2HDMs is still valid. }. 

As we discussed in Sec.~\ref{sec:2hdm_x}, the value of $\sin(\beta-\alpha)$ describes ``{\it SM-like ness}'' of $h$, namely, all the $h$ couplings to the SM particles 
become the same as those in the SM prediction in the limit of $\sin(\beta-\alpha)\to 1$. 
In other words, once $\sin(\beta-\alpha)\neq 1$ is given, both the $hVV$ and $hf\bar{f}$ couplings deviate from those of the SM values. 
In our scenario, the value of $\sin(\beta-\alpha)$ is determined from Eq.~(\ref{scenario}). 
Thus, a small but non-zero deviation from the SM-like limit is given.

\begin{figure}[t]
\centering
\includegraphics[width=0.4\hsize]{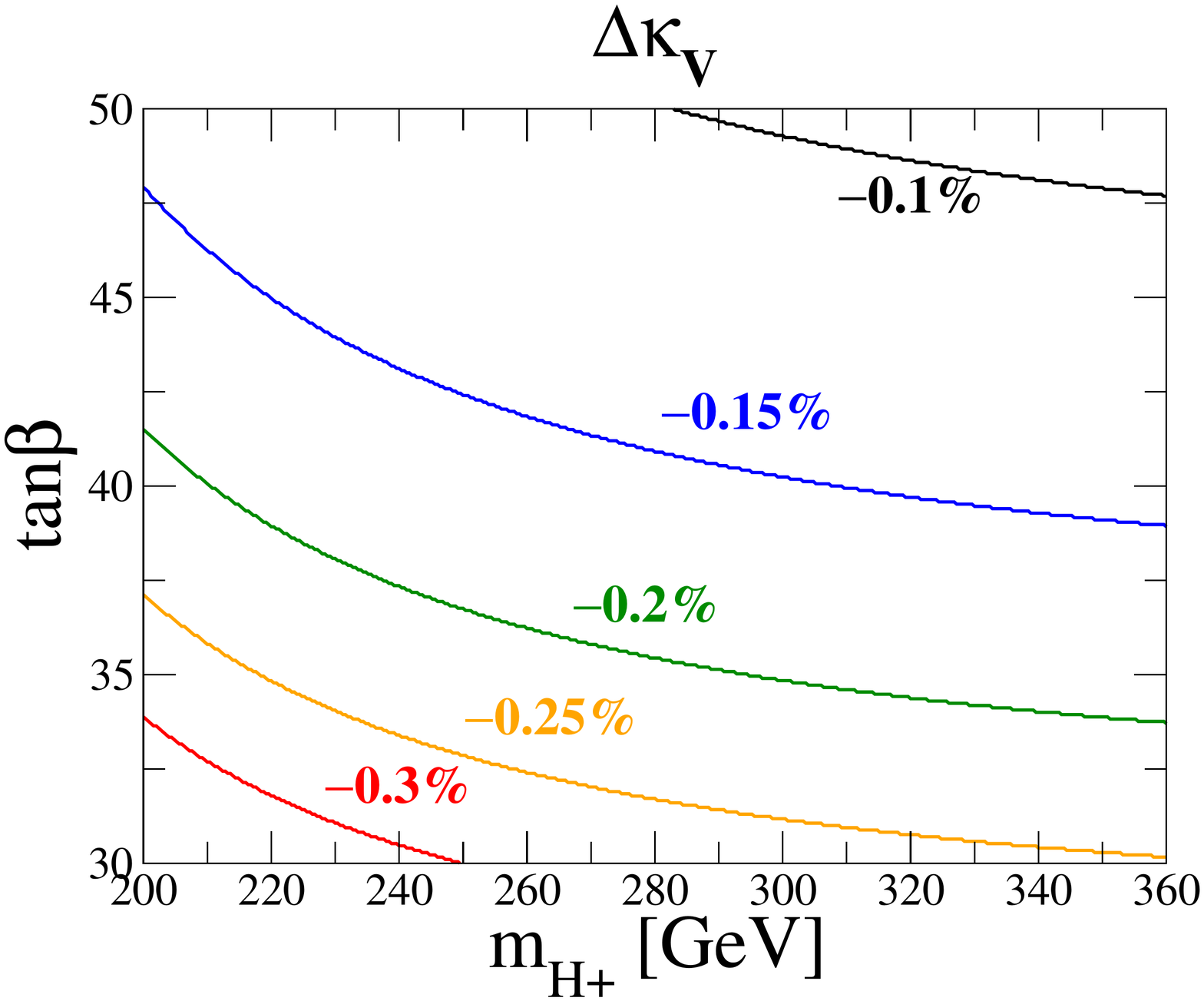} \hspace{-5mm}
\includegraphics[width=0.4\hsize]{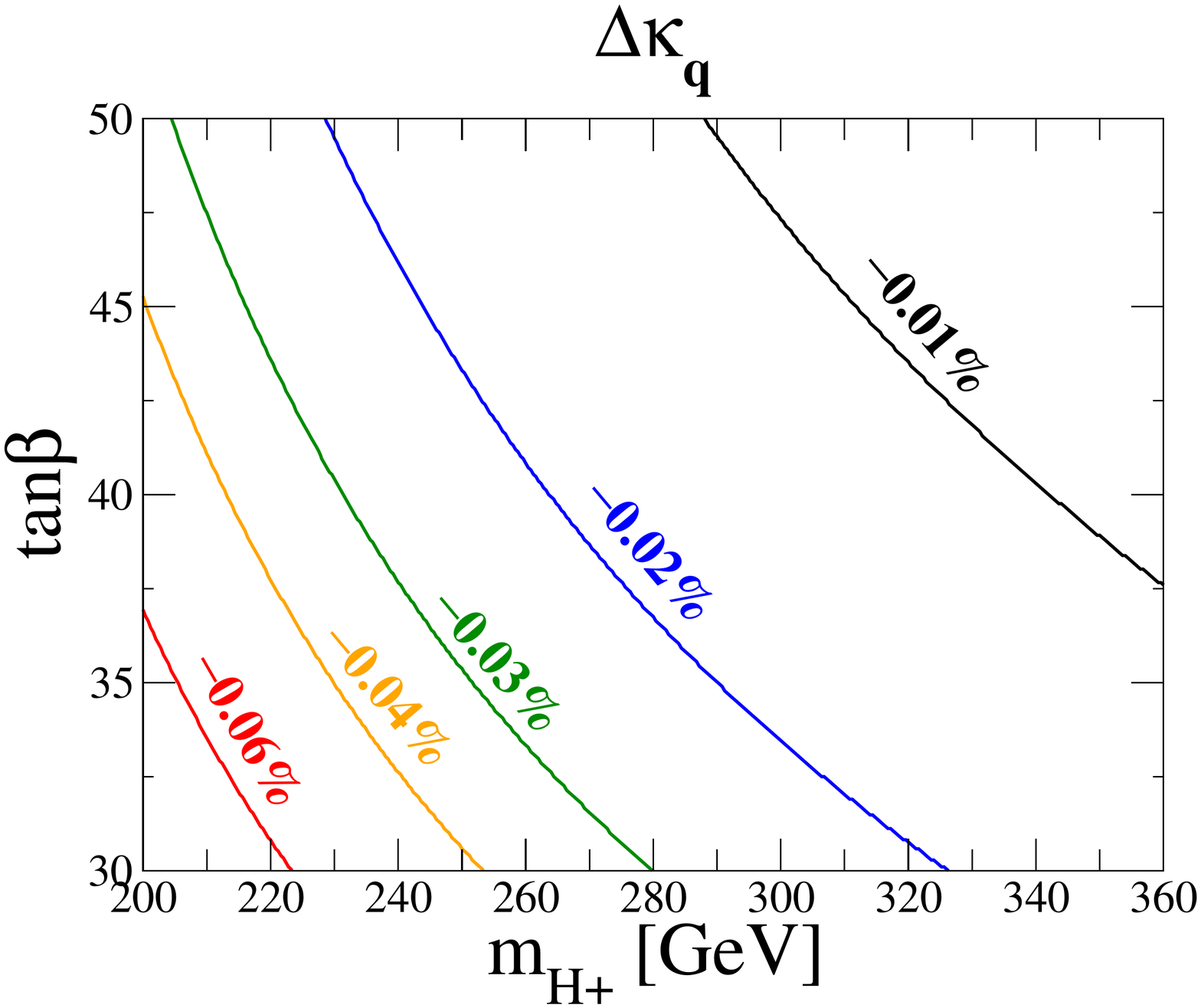}  \\
\includegraphics[width=0.4\hsize]{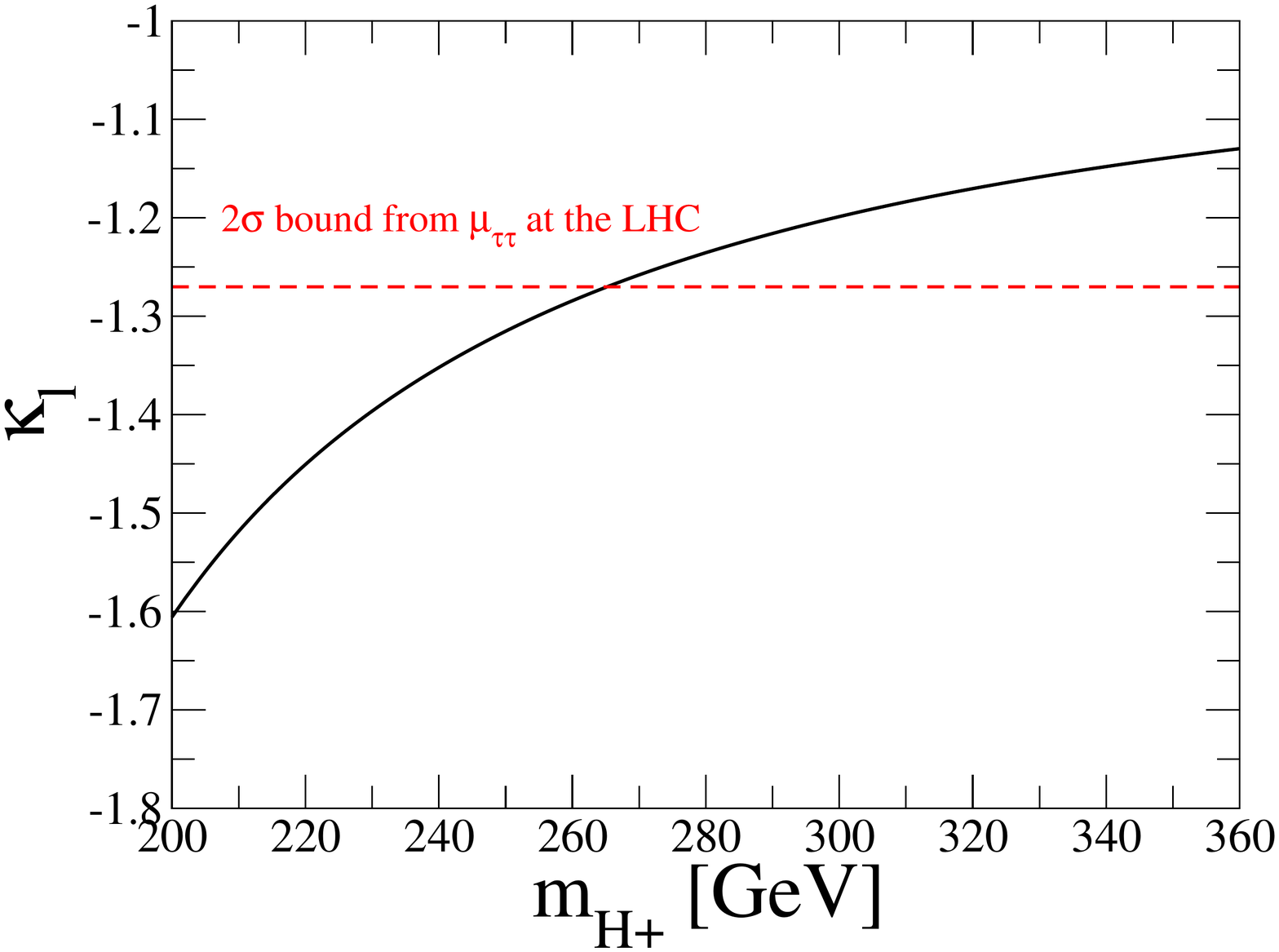}
\caption{Contour plots for $\Delta\kappa_V^{}$ (upper left) and  
$\Delta\kappa_q^{}$  (upper right) 
on the $m_{H^\pm}$-$\tan\beta$ plane, where $\Delta\kappa_X=\kappa_X-1$. 
The $m_{H^\pm}$ dependence of $\Delta\kappa_\ell^{}$ is shown in the lower panel with $\tan\beta=35$. 
We take $m_H^{}=M=m_{H^\pm}$ and $m_A^{}=20$ GeV in all the panels.  
The horizontal dashed line represents the bound from the signal strength using Eq.~(\ref{strength}). }\label{fig:kappa}
\end{figure}

In order to describe the deviation in the $h$ couplings, 
we introduce the so-called scaling factors defined as $\kappa_X^{}=g_{hXX}/g_{hXX}^{\rm SM}$ and its deviation from unity; i.e, 
$\Delta\kappa_X^{}=\kappa_X^{}-1$. 
From Eqs.~(\ref{xih}) and (\ref{hvv}) and the approximate formulae given in Eqs.~(\ref{sin_app}) and (\ref{cos_app}), 
we obtain 
\begin{align}
&\Delta\kappa_V^{} \simeq -\frac{2}{\tan^2\beta}\left(1+\frac{m_h^2}{m_{H^\pm}^2}-\frac{2m_A^2}{m_{H^\pm}^2}  \right), \label{delkv}\\
&\Delta\kappa_q^{} \simeq -\frac{2}{\tan^2\beta}\left(\frac{m_h^2}{2m_{H^\pm}^2}-\frac{m_A^2}{m_{H^\pm}^2}  \right), \label{delkq}\\
&\kappa_\ell^{} \simeq -1 -\frac{m_h^2}{m_{H^\pm}^2}+\frac{2m_A^2}{m_{H^\pm}^2}. \label{delkl}
\end{align}

In the upper panels of Fig.~\ref{fig:kappa}, we show the contour plots for $\Delta\kappa_V$ and $\Delta\kappa_q$, where $\Delta\kappa_X=\kappa_X-1$, 
on the $m_{H^\pm}$-$\tan\beta$ plane. 
In the lower panel, 
we show the $m_{H^\pm}$ dependence of $\kappa_\ell$ instead of showing contour plots, 
because the $\tan\beta$ dependence of $\kappa_\ell$ can be neglected as seen in Eq.~(\ref{delkl}). 
For definiteness, we take $\tan\beta=35$ in the plot for $m_{H^\pm}$-$\kappa_\ell$.
We find that the deviations in the $hVV$ and $hq\bar{q}$ couplings are
respectively $-\mathcal{O}(0.1)\%$ and $-\mathcal{O}(0.01)\%$ 
which can also be estimated from Eqs.~(\ref{delkv}), (\ref{delkq}). 
For the $h\ell\ell$ coupling, we find that its magnitude is maximally about 1.6 times larger than the SM prediction, and its sign is opposite to the SM one~\cite{Wang:2014sda}.   
From the measurement of the signal strength of the $h\to\t\t$ channel, $i.e.$, $\mu_{\tau\tau}$ at the LHC, 
the magnitude of $\kappa_\ell^{}$ is constrained. 
The definition of the signal strength is given as 
\begin{align}
\m_{XY}^{} \equiv \frac{\sigma_h \times {\rm Br}(h\to XY)}{[\sigma_h\times {\rm Br}(h\to XY)]_{\rm SM}}, 
\end{align}
where 
$\sigma_h$ and Br($h\to XY$) are respectively the production cross section of the SM-like Higgs boson $h$ and the decay branching fraction of the $h\to XY$ mode. 
In our parameter set, $\sigma_h$ is almost the same as that in the SM, because of the small $\Delta\kappa_q$ and  $\Delta\kappa_V^{}$, so that 
the signal strength is simply given as the ratio of the branching fraction as 
$\mu_{XY}^{}\simeq {\rm Br}(h\to XY)/{\rm Br}(h\to XY)_{\rm SM}$.

The ATLAS and CMS collaborations report the signal strength as
$\m_{\t\t}=1.43^{+0.43}_{-0.37}$ \cite{Aad:2015vsa} and
$\m_{\t\t}=0.91 \pm 0.28$ \cite{CMS:2014ega}, respectively.
By taking a naive combination of them\footnote{
To obtain the naive combination, we treat the signal strength from ATLAS as $1.43 \pm 0.40$ by taking the average of errors.}, 
we obtain
\begin{align}
\m_{\t\t} = 1.08 \pm 0.23. \label{strength}
\end{align}
Thus, the region with $|\k_\ell| > 1.27$ is excluded at $2\s$ level, which corresponds to the constraint of 
$m_{H^\pm} \lesssim 270~{\rm GeV}$ as seen in Fig.~\ref{fig:kappa}.

\begin{figure}[p]
\centering
\includegraphics[width=0.52\hsize]{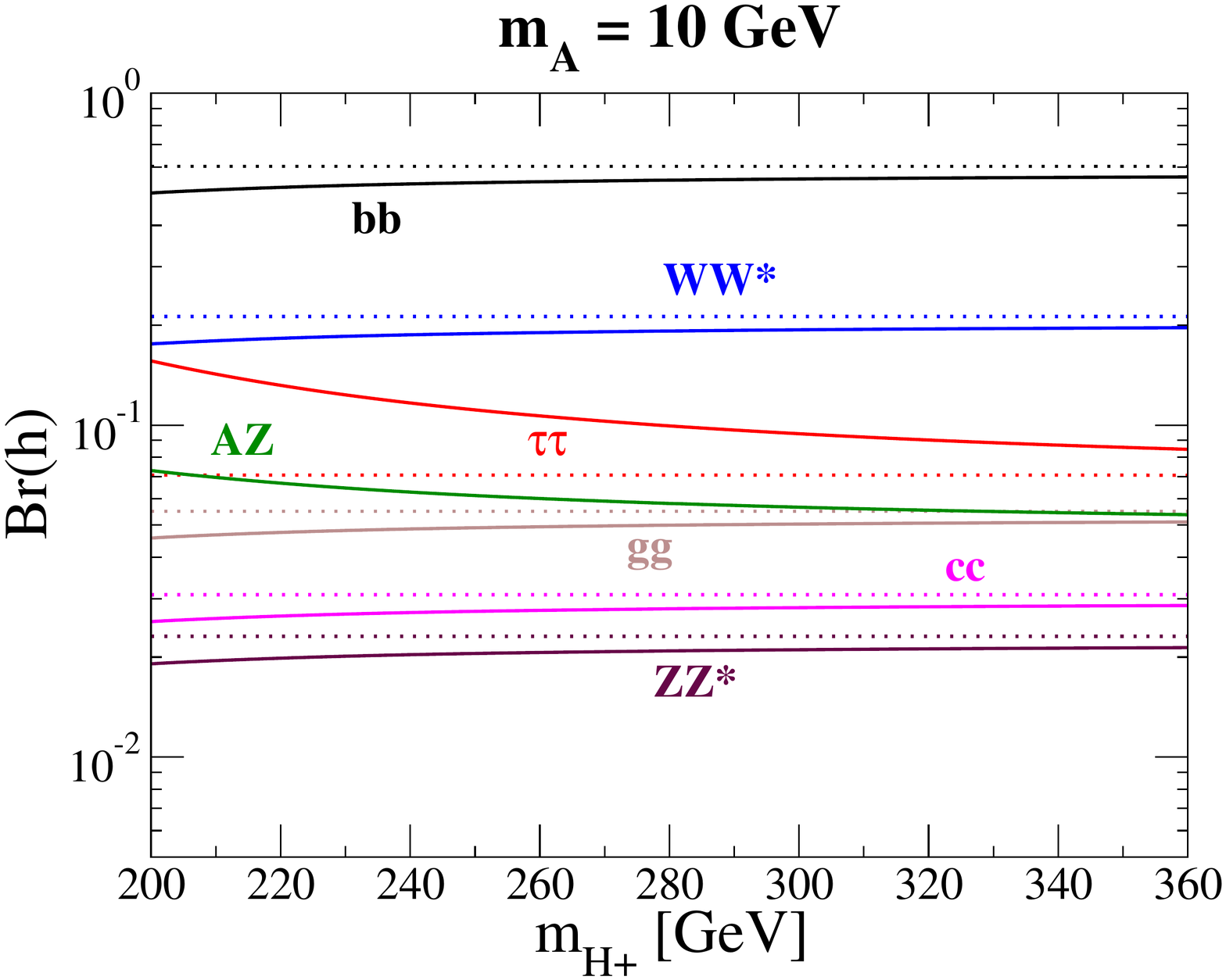}\\
\includegraphics[width=0.52\hsize]{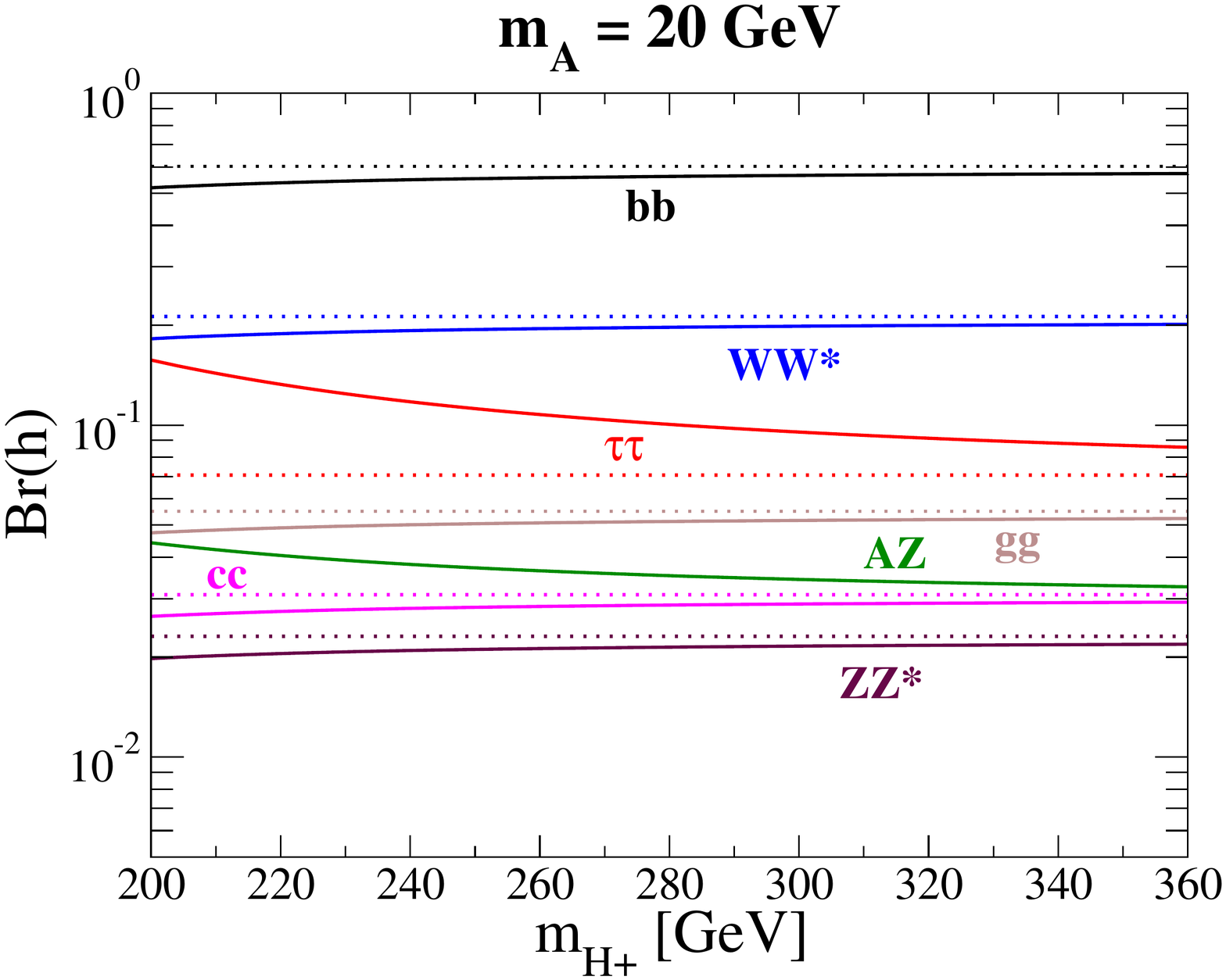}\\
\includegraphics[width=0.52\hsize]{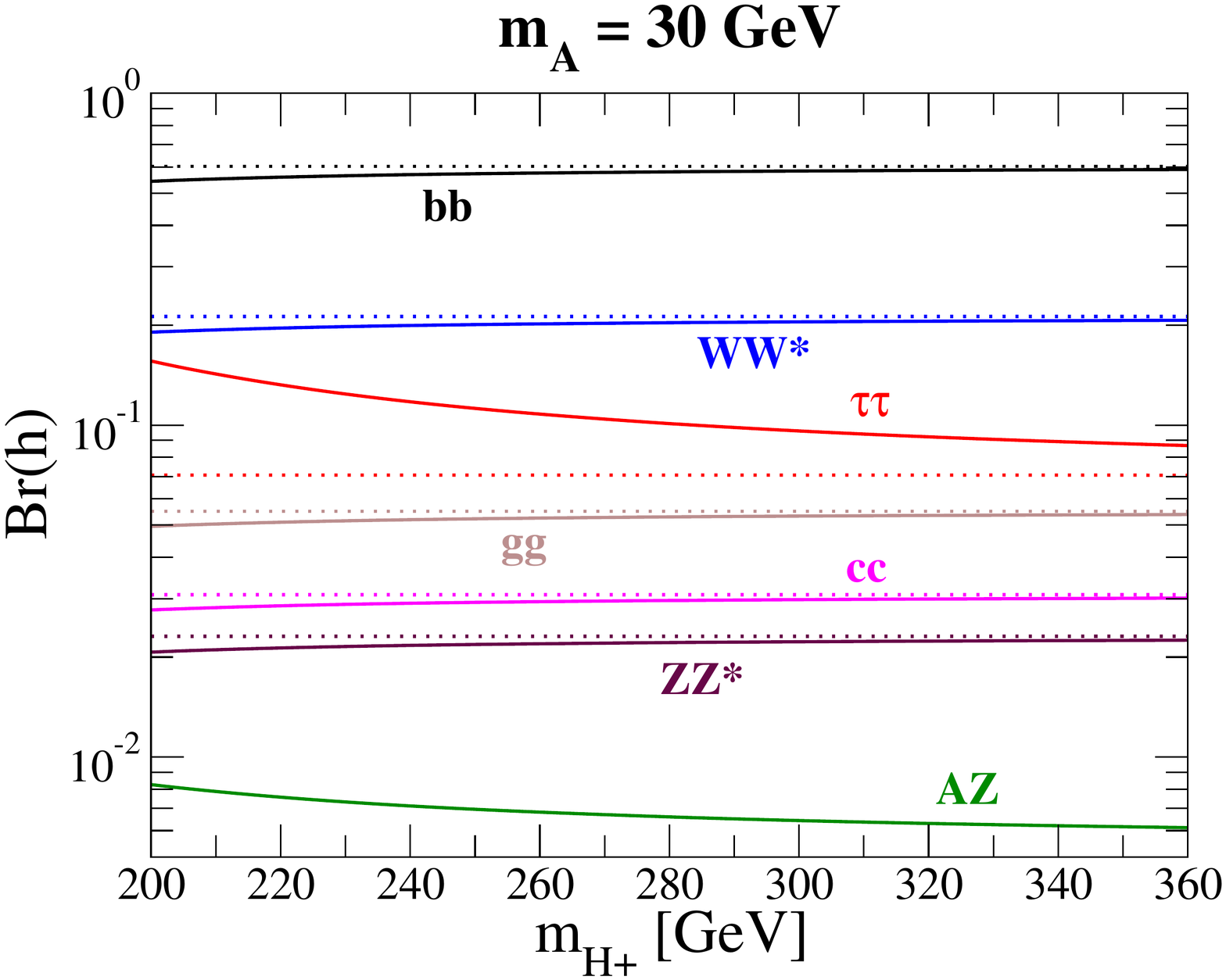}  
\caption{The branching fraction of $h$ as a function of $m_{H^\pm}^{}$ in the case of $m_H^{}=M=m_{H^\pm}$ and $\tan\beta=35$.  
The top, middle and bottom panels show the cases with $m_A^{}=10$, 20 and 30 GeV, respectively. 
The solid (dotted) curves shows the prediction in the Type-X 2HDM (SM). }\label{fig:hdecay}
\end{figure}

The deviation in the $h$ couplings makes a difference in the branching fraction of $h$ from the SM prediction. 
In Fig.~\ref{fig:hdecay}, we show the branching fraction of $h$ as a function of $m_{H^\pm}$. 
Because only the magnitude of the $h\ell\ell$ coupling can be larger than the SM prediction, 
only the branching fraction of the $h\to \tau\tau$ mode is enhanced. 
On the other hand, that of all the other modes shown in this figure are reduced. 
The size of deviations becomes smaller as $m_{H^\pm}$ increases.
We note that the branching fraction of the $h\to \mu\mu$ decay is also enhanced similar to the $h\to\tau\tau$ mode. 
When $m_{H^\pm}=300$ GeV, and all the other parameters are taken to be the same as in Fig.~\ref{fig:hdecay},
Br$(h\to \mu\mu)$ is about $3.4~(2.5)\times 10^{-4}$ in the Type-X 2HDM (SM).

It is important to mention here that there appears
the exotic decay mode $h \to AZ$ in our parameter set as seen in Fig.~\ref{fig:hdecay}. 
Although the coupling constant of the $hZA$ interaction 
is proportional to $\cos(\b-\a)$ which is suppressed by $\cot\beta$ as seen in Eq.~(\ref{cos_app}), 
its branching fraction is not so small, especially for the case with small $m_A$. For example,  we obtain
Br($h\to AZ)\simeq 7\%$ with $m_A^{}=10$ GeV and $m_{H^\pm}=200$ GeV. 
This new decay mode of $h$
gives the additional contribution to the four-lepton channel in the SM Higgs boson search 
if $A$ decays into a pair of muon. 
In the present scenario, ${\rm Br}(A\to\m\m) \simeq {\rm Br}(A\to\t\t)\times (m_\m / m_\t)^2 \simeq 0.0036$ is obtained. 
On the other hand, from Higgs boson searches at the LHC, the upper bound on ${\rm Br}(h\to ZA) \times {\rm Br}(A\to \ell\ell)$ 
with $\ell=e$ or $\mu$ is given as $5\text{-}10 \times 10^{-4}$ for $12<m_A<34~{\rm GeV}$ \cite{Curtin:2013fra}.
This limit is translated into the bound ${\rm Br}(h\to AZ)\lesssim 14\text{-}28\%$ in the Type-X 2HDM. 
The typical size of Br($h\to AZ)$ is below the upper bound as explained in the above. 
In addition to this channel, $e$ and $\m$ are produced from the leptonic decay of $\t$.
Thus, the $ZA \to \ell\ell \t\t \to 4\ell+E_T\hspace{-4.5mm}/$\hspace{2mm} channel can also contribute to the four lepton channel
even though the invariant mass distribution of the four lepton system is different from that by $ZZ^* \to 4\ell$.
This will be a subject of a future work.

\begin{figure}[t]
\centering
\includegraphics[width=0.8\hsize]{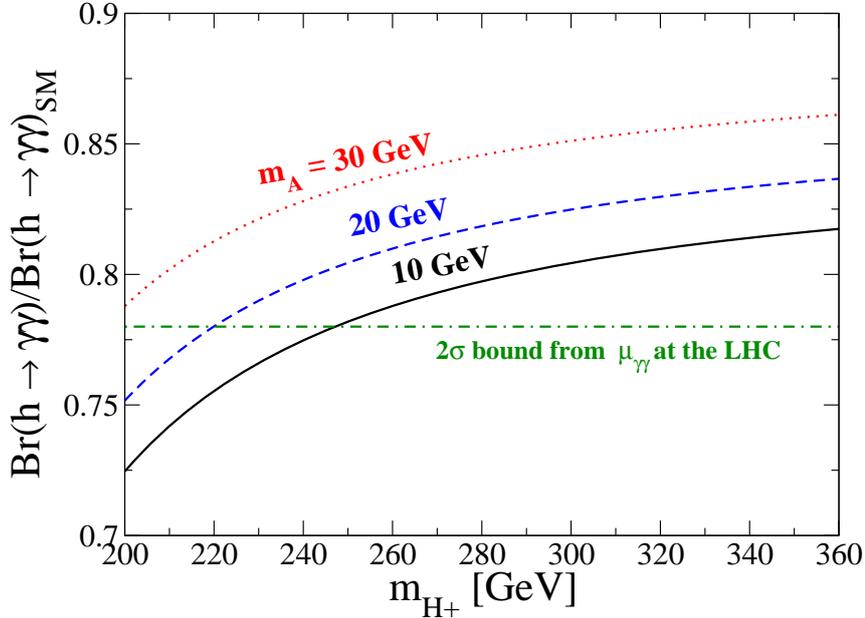}
\caption{Ratio of the branching fraction $\text{Br}(h\to\g\g) / \text{Br}(h\to\g\g)_{\rm SM}$ in our scenario with $\tan\beta=35$.
The solid, dashed and dotted curves show the cases with $m_A$=10, 20 and 30 GeV, respectively.  
The horizontal dashed line shows the bound from $\mu_{\gamma\gamma}$ given in Eq.~(\ref{mu_gamgam_exp}) at 2$\sigma$ level. 
}\label{fig:diphoton_plot}
\end{figure}

Next, we discuss the one-loop induced $h\to \gamma\gamma$ decay mode. 
Because of the $H^\pm$ contribution, the decay rate can be significantly modified even if 
the $h$ couplings are not changed so much from the SM prediction.
We note that the deviation in the $h\ell\ell$ coupling can be neglected in the decay rate of the $h\to \gamma\gamma$ mode, 
because its effect appears in the tau loop contribution, but the tau Yukawa coupling is too small as compared to the top Yukawa coupling.  
The decay rate of the diphoton mode is given as
\begin{align}
&\G(h\to \g\g)
= \frac{G_F \a_{\text{em}}^2 m_h^3}{128\sqrt{2} \pi^3}\left|
s_{\beta-\alpha}A_{1}(\tau_W^{}) + \sum_f \xi_f^hA_{1/2}^f(\tau_f^{}) + \frac{\l_3 v^2}{2m_{H^\pm}^2} A_0(\tau_{H^\pm}^{}) 
\right|^2,\notag\\
&~~\text{with}~~\tau_X^{}= \frac{m_h^2}{4m_X^2} \label{eq:diphoton_s2},
\end{align}
where the $A_{1}$, $A_{1/2}^f$ and $A_0$ terms correspond to the $W$ boson, fermion and $H^\pm$ loop contributions, respectively. 
Each of the loop functions are given by 
\begin{align}
A_1(\t) &= -\t^{-2}[ 2\t^2+3\t+3(2\t-1)f(\t) ], \\
A_{1/2}^f(\t)&= 2N_c^f Q_f^2 \t^{-2}[\t + (\t-1)f(\t)], \\
A_0(\t) &= -\t^{-2}[ \t-f(\t) ],
\end{align}
where $f(\t)$ is defined as
\begin{align}
f(\t) = \left\{
\begin{array}{ll}
\arcsin^2\sqrt{\t}. & (\t \leq 1) \\
-\displaystyle\frac{1}{4}\left(\log\displaystyle\frac{1+\sqrt{1-\t^{-1}}}{1-\sqrt{1-\t^{-1}}} - i\pi \right)^2. & (\t > 1)
\end{array}
\right.
\end{align}
We note that in the SM-like limit, which gives a good approximation for the numerical study in our scenario, 
we obtain the numerical value of the SM $W$ and top loop contributions; \textit{i.e.}, 
$A_{1}(\tau_W^{}) + A_{1/2}^t(\tau_t^{})\simeq -6.45$. 

We here discuss the impact of the $H^\pm$ loop contribution to the $h\to \gamma\gamma$ decay. 
In our scenario, since the light CP-odd Higgs boson is required to explain the muon $g-2$ anomaly, 
the relatively large mass difference between $A$ and $H^\pm$ is needed to avoid the current experimental constraints, 
which is generated by the Higgs quartic couplings.
This can be clearly seen by rewriting the $\lambda_3$ coupling appearing in Eq.~(\ref{eq:diphoton_s2}) as follows
\begin{align}
\l_3 v^2 = 2(m_{H^\pm}^2 - m_{A}^2) + \l_{hAA} v^2, 
\end{align}
where the second term is now zero according to our benchmark scenario shown in Eq.~(\ref{scenario}). 
In addition, the loop function $A_0$ is expanded as $A_0(x) = 1/3 + 8x/45 + \mathcal{O}(x^2)$ under $x\ll 1$ or equivalently $m_{H^\pm} \gg m_h$. 
Thus, even if $H^\pm$ is relatively heavy, its contribution is not decoupled as we can see  
the asymptotic behavior $\lambda_3 v^2 /(2 m_{H^\pm}^2) A_0(\tau_{H^\pm}) \to 1/3$   $(m_{H^\pm} \to \infty)$.

The ATLAS and CMS collaborations report
the signal strength as
$\m_{\g\g}=1.17 \pm 0.27$ \cite{Aad:2014eha} and
and $\m_{\g\g}=1.12 \pm 0.24$ \cite{CMS:2014ega}, respectively.
By taking naive combination of them, we obtain
\begin{align}
\m_{\g\g} = 1.14 \pm 0.18. \label{mu_gamgam_exp}
\end{align}

In Fig.~\ref{fig:diphoton_plot}, we show the ratio of the branching fraction of the $h\to \gamma\gamma$ mode as a function of $m_{H^\pm}$ with $\tan\beta=35$. 
We find that the deviation in the branching fraction from the SM prediction is obtained in the range of $-30$\% to $-15\%$.
The expected accuracy for the measurement of the decay rate of the diphoton mode 
is around $-10$\% at the LHC 14 TeV 300 fb$^{-1}$ and 5\% at the ILC \cite{Peskin:2012we}.
Therefore, our scenario is also probed by detecting the deviation in the $h\to \g\g$ decay rate in addition to the extra Higgs boson searches discussed in 
Sec.~\ref{sec:extra_Higgs} and the measurement of $\kappa_\ell$.
By looking at the horizontal line representing the bound from $\mu_{\gamma\gamma}$, 
we see that $m_{H^\pm}\lesssim 220$ GeV is excluded in the case of $m_A^{}=20$ GeV, which weaker than the constraint of $\k_\ell$.

\section{Conclusion}\label{sec:conclusion}

In this paper, we have explored the possibility to explain the muon $g-2$ anomaly in the Type-X 2HDM.
We have shown in Fig.~\ref{fig:typeX} that
the measurement of the leptonic tau decay gives an important constraint on the parameter space.
As a result, the region which can explain the discrepancy in the muon $g-2$ at the $1\sigma$ level 
is excluded by the constraint from the tau decay, and that at the $2\sigma$ level is allowed. 
We have found that the parameter space with 
$10 \lsim m_A \lsim 30~{\rm GeV}$, 
$200 \lsim m_{H,H^\pm} \lsim 350~{\rm GeV}$
and $30 \lsim \tan\b \lesssim 50$
is favored by the explanation of the anomaly at the $2\sigma$ level. 

After finding the viable parameter region for the muon $g-2$,
we have discussed the implication of the favored parameter region to the collider phenomenology. 
In our scenario, 
the $4\tau$, $3\tau$, $4\tau+Z$ and $4\tau+W$ signatures are expected from the electroweak productions of the extra Higgs bosons at the LHC. 
The cross sections of these signals are shown in Table~\ref{tab_xs}. 

Finally, we have investigated the possible deviation in the property of the SM-like Higgs boson $h$. 
We have found that the value of the $h\ell\ell$ coupling is predicted to be the SM prediction times about $-1.6$ to $-1.0$,
and the current data of the signal strength $\mu_{\tau\tau}$ at the LHC 
excludes $m_{H^\pm} \lesssim 270~{\rm GeV}$ in the case of $m_{H^\pm} = m_H$ and $m_A=20~{\rm GeV}$.
We also have evaluated the branching fraction 
of the $h\to \g\g$ mode which is modified by the non-decoupling charged Higgs boson loop effect. 
We have shown that the deviation in the branching fraction of the $h\to \gamma\gamma$ mode is about $-30\%$ to $-15\%$ which
can be detected at the future collider experiments such as the high-luminosity LHC and the ILC. 
Therefore, the precise measurements of the branching fractions of $h\to \g\g$  and $h\to\t\t$
gives an important indirect test to probe our scenario.
Furthermore, we have found that the branching fraction of $h\to ZA$ becomes a few percent, and it 
also would provide another important signature to test our scenario.
Therefore, 
in the Type-X 2HDM motivated by the explanation of the muon $g-2$ anomaly, 
there are various characteristic deviations in the 125 GeV Higgs boson property, which 
will be tested at collider experiments in the near future.

\section*{Acknowledgements}
We used 
Feynmf~\cite{Ohl:1995kr} to draw the Feynman diagrams.
The work is supported by Grant-in-Aid for Scientific research from the Ministry of Education, Culture, Sports, Science and Technology (MEXT), Japan, No.~23104006 [TA],
JSPS Research Fellowships for Young Scientists [RS]
and JSPS Postdoctoral Fellowships for Research Abroad [KY].
\appendix

\section{Decay rates for the extra Higgs bosons}\label{sec:yyy}

The decay rates of the extra Higgs bosons 
into the fermion pair are calculated at the tree level as 
\begin{align}
\Gamma(H^\pm \to f\bar{f}')&=\sqrt{2}G_F\frac{m_{H^\pm}^3}{8\pi} N_c^f|V_{ff'}|^2\lambda^{1/2}(x_{f H^\pm }^{},x_{f' H^\pm }^{})\notag\\
& \times\left[(x_{f H^\pm }^{}\xi_f^2+x_{f' H^\pm }^{}\xi_{f'}^2)(1-x_{fH^\pm }^{}-x_{f'H^\pm }^{})+4x_{fH^\pm }^{}x_{f'H^\pm }^{}\xi_f\xi_{f'}\right], \\
\Gamma(H\to f\bar{f})&=
\sqrt{2}G_F\frac{m_H^{} m_f^2}{8\pi}(\xi_f^H)^2N_c^f\lambda^{3/2}\left(x_{fH }^{},x_{fH}^{}\right), \\
\Gamma(A\to f\bar{f})&=
\sqrt{2}G_F\frac{m_A m_f^2}{8\pi}\xi_f^2N_c^f\lambda^{1/2}\left(x_{fA}^{},x_{fA}^{}\right), 
\label{decay_rates}
\end{align}
where
$x_{ab}=m_a^2/m_b^2$, $N_c^f$ is the color factor, and the two body phase space function $\lambda(x,y)$ is given by 
\begin{align}
\lambda(x,y) = 1+x^2+y^2-2x-2y-2xy. 
\end{align}
We note that the above formulae are applied to all the types of Yukawa interactions. 
The decay rates into the gauge boson associated modes are calculated at the tree level as 
\begin{align}
\Gamma(H\to AZ) &= s_{\beta-\alpha}^2\frac{\sqrt{2}G_F}{16\pi }m_H^3\lambda^{3/2}\left(x_{ZH}^{},x_{AH}^{}\right), \\
\Gamma(H^\pm \to A W^\pm ) &= \frac{\sqrt{2}G_F}{16\pi }m_{H^\pm}^3\lambda^{3/2}\left(x_{WH^\pm}^{},x_{AH^\pm}^{}\right). 
\end{align}

\section{Electroweak production of a pair of Higgs bosons}\label{sec:xsec}
Spin and color-averaged parton level cross sections are given as \cite{Djouadi:2005gj}
\begin{align}
\hat\s(\hat s; u\bar d\to H^+ A) &=
\frac{G_F^2 m_W^4}{72\pi} \frac{s^2_{\b-\a}}{\hat s} \left( \frac{\hat s}{\hat s-m_W^2} \right)^2 \l^{3/2}\left(\frac{m_{H^\pm}^2}{\hat s},\frac{m_A^2}{\hat s}\right), \label{c1}\\
\hat\s(\hat s; u\bar d\to H^+ H) &=
\frac{G_F^2 m_W^4}{72\pi} \frac{s^2_{\b-\a}}{\hat s} \left( \frac{\hat s}{\hat s-m_W^2} \right)^2\l^{3/2}\left(\frac{m_{H^\pm}^2}{\hat s},\frac{m_H^2}{\hat s}\right), \label{c2}\\
\hat\s(\hat s; q\bar q\to H A) &=
\frac{G_F^2 m_Z^4}{18 \pi} (v_q^2 + a_q^2) \frac{s^2_{\b-\a}}{\hat s} \left( \frac{\hat s}{\hat s-m_Z^2} \right)^2 \l^{3/2}\left(\frac{m_{H}^2}{\hat s},\frac{m_A^2}{\hat s}\right),\\
\hat\s(\hat s; q\bar q\to H^+ H^-) &=
\frac{\pi \a_{\text{em}}^2}{9\hat s}\left[ \left(Q_q - \frac{v_q v_H}{s_W^2 c_W^2} \frac{\hat s}{\hat s-m_Z^2} \right)^2 + \frac{a_q^2 v_H^2}{s_W^4 c_W^4} \left(\frac{\hat s}{\hat s-m_Z^2}\right)^2  \right] \notag\\
&\times \l^{3/2}\left(\frac{m_{H^\pm}^2}{\hat s},\frac{m_{H^\pm}^2}{\hat s}\right),
\end{align}
where $v_H = -1/2+s_W^2$, and $v_q$ and $a_q$ are defined in Eq.~(\ref{vector_coupling}).
Production cross section at the hadron collider is calculated as
\begin{align}
\s(pp\to AB) &= 2\int_0^1 d\t \sum_{q_1, \bar q_2} \frac{d{\cal L}_{q_1 \bar q_2}}{d\t} \hat\s(q_1 \bar q_2 \to A B ;\hat s = \t s).\label{lumi}
\end{align}
where ${\cal L}$ is the parton luminosity which is defined as
\begin{align}
\frac{d {\cal L}_{q_1 \bar q_2} }{d\t} &= \int_\t^1 \frac{dx}{x} q_1(x) \bar q_2(\t/x).\label{lumi2}
\end{align}
We note that the parton level cross sections of 
$\bar ud\to H^- A$ and  $\bar ud\to H^- H$ processes are the same as those given in Eqs.~(\ref{c1}) and (\ref{c2}), respectively. 
However, because of the difference of parton luminosity of the initial proton, 
the cross section of $H^+H/A$ and $H^-H/A$ are different in the stage after the convolution of the luminosity function shown in Eqs.~(\ref{lumi}) and (\ref{lumi2}).


\begin{thebibliography}{99}


\bibitem{Bennett:2006fi} 
  G.~W.~Bennett {\it et al.}  [Muon G-2 Collaboration],
  Phys.\ Rev.\ D {\bf 73}, 072003 (2006)
  [hep-ex/0602035].

\bibitem{Davier:2010nc} 
  M.~Davier, A.~Hoecker, B.~Malaescu and Z.~Zhang,
  Eur.\ Phys.\ J.\ C {\bf 71}, 1515 (2011)
  [Erratum-ibid.\ C {\bf 72}, 1874 (2012)]
  [arXiv:1010.4180 [hep-ph]].

\bibitem{Hagiwara:2011af} 
  K.~Hagiwara, R.~Liao, A.~D.~Martin, D.~Nomura and T.~Teubner,
  J.\ Phys.\ G {\bf 38}, 085003 (2011)
  [arXiv:1105.3149 [hep-ph]].

\bibitem{Grange:2015fou} 
  J.~Grange {\it et al.}  [Muon g-2 Collaboration],
  arXiv:1501.06858 [physics.ins-det].

\bibitem{Iinuma:2011zz} 
  H.~Iinuma [J-PARC New g-2/EDM experiment Collaboration],
  J.\ Phys.\ Conf.\ Ser.\  {\bf 295}, 012032 (2011). 


\bibitem{Czarnecki:2002nt} 
  A.~Czarnecki, W.~J.~Marciano and A.~Vainshtein,
  Phys.\ Rev.\ D {\bf 67}, 073006 (2003)
  [Erratum-ibid.\ D {\bf 73}, 119901 (2006)]
  [hep-ph/0212229].

\bibitem{Jegerlehner:2009ry} 
  F.~Jegerlehner and A.~Nyffeler,
  Phys.\ Rept.\  {\bf 477}, 1 (2009)
  [arXiv:0902.3360 [hep-ph]].

\bibitem{GW}
  S.~L.~Glashow and S.~Weinberg,
  Phys.\ Rev.\  D {\bf 15}, 1958 (1977).

\bibitem{Barger}
  V.~D.~Barger, J.~L.~Hewett and R.~J.~N.~Phillips,
  Phys.\ Rev.\ D {\bf 41}, 3421 (1990). 

\bibitem{Grossman}
  Y.~Grossman,
  Nucl.\ Phys.\ B {\bf 426}, 355 (1994).


\bibitem{Aoki:2009ha} 
  M.~Aoki, S.~Kanemura, K.~Tsumura and K.~Yagyu,
  Phys.\ Rev.\ D {\bf 80}, 015017 (2009)
  [arXiv:0902.4665 [hep-ph]].


\bibitem{Dedes:2001nx} 
  A.~Dedes and H.~E.~Haber,
  JHEP {\bf 0105}, 006 (2001)
  [hep-ph/0102297]. 


\bibitem{Chang:2000ii} 
  D.~Chang, W.~F.~Chang, C.~H.~Chou and W.~Y.~Keung,
  Phys.\ Rev.\ D {\bf 63}, 091301 (2001)
  [hep-ph/0009292]. 

\bibitem{Kingman1}
  K.~m.~Cheung, C.~H.~Chou and O.~C.~W.~Kong,
  Phys.\ Rev.\ D {\bf 64}, 111301 (2001)
  [hep-ph/0103183].


\bibitem{Krawczyk:2001pe} 
  M.~Krawczyk,
  hep-ph/0103223.

\bibitem{Kingman2} 
  K.~Cheung and O.~C.~W.~Kong,
  Phys.\ Rev.\ D {\bf 68}, 053003 (2003)
  [hep-ph/0302111].

\bibitem{Wu:2001vq} 
  Y.~L.~Wu and Y.~F.~Zhou,
  Phys.\ Rev.\ D {\bf 64}, 115018 (2001)
  [hep-ph/0104056].

\bibitem{Gunion:2008dg} 
  J.~F.~Gunion,
  JHEP {\bf 0908}, 032 (2009)
  [arXiv:0808.2509 [hep-ph]].


 
\bibitem{Cao:2009as} 
  J.~Cao, P.~Wan, L.~Wu and J.~M.~Yang,
  Phys.\ Rev.\ D {\bf 80}, 071701 (2009)
  [arXiv:0909.5148 [hep-ph]].

\bibitem{Chun} 
  A.~Broggio, E.~J.~Chun, M.~Passera, K.~M.~Patel and S.~K.~Vempati,
  arXiv:1409.3199 [hep-ph].

\bibitem{Wang:2014sda} 
  L.~Wang and X.~F.~Han,
  arXiv:1412.4874 [hep-ph].

\bibitem{Ilisie:2015tra} 
  V.~Ilisie,
  arXiv:1502.04199 [hep-ph].

\bibitem{Barr-Zee1} 
  J.~D.~Bjorken and S.~Weinberg,
  Phys.\ Rev.\ Lett.\  {\bf 38}, 622 (1977). 

\bibitem{Barr-Zee2}
  S.~M.~Barr and A.~Zee,
  Phys.\ Rev.\ Lett.\  {\bf 65}, 21 (1990)
  [Erratum-ibid.\  {\bf 65}, 2920 (1990)].

\bibitem{Aad:2012tfa} 
  G.~Aad {\it et al.}  [ATLAS Collaboration],
  Phys.\ Lett.\ B {\bf 716}, 1 (2012)
  [arXiv:1207.7214 [hep-ex]].

\bibitem{Chatrchyan:2012ufa} 
  S.~Chatrchyan {\it et al.}  [CMS Collaboration],
  Phys.\ Lett.\ B {\bf 716}, 30 (2012)
  [arXiv:1207.7235 [hep-ex]].


\bibitem{CMSandLHCbCollaborations:2013pla} 
  CMS and LHCb Collaborations [CMS and LHCb Collaboration],
  CMS-PAS-BPH-13-007.





\bibitem{Ma:2002pf} 
  E.~Ma and D.~P.~Roy,
  Nucl.\ Phys.\ B {\bf 644}, 290 (2002)
  [hep-ph/0206150].


\bibitem{Aoki:2008av} 
  M.~Aoki, S.~Kanemura and O.~Seto,
  Phys.\ Rev.\ Lett.\  {\bf 102}, 051805 (2009)
  [arXiv:0807.0361 [hep-ph]].





\bibitem{Hollik} 
  W.~Hollik and T.~Sack,
  Phys.\ Lett.\ B {\bf 284}, 427 (1992). 

\bibitem{Krawczyk:2004na} 
  M.~Krawczyk and D.~Temes,
  Eur.\ Phys.\ J.\ C {\bf 44}, 435 (2005)
  [hep-ph/0410248].

\bibitem{Logan:2009uf} 
  H.~E.~Logan and D.~MacLennan,
  Phys.\ Rev.\ D {\bf 79}, 115022 (2009)
  [arXiv:0903.2246 [hep-ph]].

\bibitem{Uni-2hdm1}

  H.~Huffel and G.~Pocsik,
  Z.\ Phys.\  C {\bf 8}, 13 (1981);

  J.~Maalampi, J.~Sirkka and I.~Vilja,
  Phys.\ Lett.\  B {\bf 265}, 371 (1991).

\bibitem{Uni-2hdm2}

  S.~Kanemura, T.~Kubota and E.~Takasugi,
  Phys.\ Lett.\  B {\bf 313}, 155 (1993). 

\bibitem{Uni-2hdm3}  
A.~G.~Akeroyd, A.~Arhrib and E.~M.~Naimi,
  Phys.\ Lett.\  B {\bf 490}, 119 (2000). 

\bibitem{Uni-2hdm4}  

  I.~F.~Ginzburg and I.~P.~Ivanov,
  Phys.\ Rev.\ D {\bf 72}, 115010 (2005).


\bibitem{VS_THDM}

  N.~G.~Deshpande and E.~Ma,
  Phys.\ Rev.\  D {\bf 18}, 2574 (1978);

  M.~Sher,
  Phys.\ Rept.\  {\bf 179}, 273 (1989);

  S.~Nie and M.~Sher,
  Phys.\ Lett.\  B {\bf 449}, 89 (1999). 

\bibitem{VS_THDM2}
  S.~Kanemura, T.~Kasai and Y.~Okada,
  Phys.\ Lett.\  B {\bf 471}, 182 (1999).




\bibitem{Abbiendi:2013hk} 
  G.~Abbiendi {\it et al.}  [ALEPH and DELPHI and L3 and OPAL and LEP Collaborations],
  Eur.\ Phys.\ J.\ C {\bf 73}, 2463 (2013)
  [arXiv:1301.6065 [hep-ex]].

\bibitem{Schael:2006cr} 
  S.~Schael {\it et al.}  [ALEPH and DELPHI and L3 and OPAL and LEP Working Group for Higgs Boson Searches Collaborations],
  Eur.\ Phys.\ J.\ C {\bf 47}, 547 (2006)
  [hep-ex/0602042].

\bibitem{Abdallah:2004wy} 
  J.~Abdallah {\it et al.}  [DELPHI Collaboration],
  Eur.\ Phys.\ J.\ C {\bf 38}, 1 (2004)
  [hep-ex/0410017].


\bibitem{ATLAS_H+} 

  G.~Aad {\it et al.}  [ATLAS Collaboration],
  arXiv:1412.6663 [hep-ex].



\bibitem{Curtin:2013fra} 
  D.~Curtin, R.~Essig, S.~Gori, P.~Jaiswal, A.~Katz, T.~Liu, Z.~Liu and D.~McKeen {\it et al.},
  arXiv:1312.4992 [hep-ph];

{\tt http://exotichiggs.physics.sunysb.edu/}


\bibitem{Djouadi:1997yw} 
  A.~Djouadi, J.~Kalinowski and M.~Spira,
  Comput.\ Phys.\ Commun.\  {\bf 108}, 56 (1998)
  [hep-ph/9704448].





\bibitem{Peskin:1991sw} 
  M.~E.~Peskin and T.~Takeuchi,
  Phys.\ Rev.\ D {\bf 46}, 381 (1992).








\bibitem{Toussaint:1978zm} 
  D.~Toussaint,
  Phys.\ Rev.\ D {\bf 18}, 1626 (1978).

\bibitem{Bertolini:1985ia} 
  S.~Bertolini,
  Nucl.\ Phys.\ B {\bf 272}, 77 (1986).

\bibitem{Pomarol:1993mu} 
  A.~Pomarol and R.~Vega,
  Nucl.\ Phys.\ B {\bf 413}, 3 (1994)
  [hep-ph/9305272].

\bibitem{Peskin:2001rw} 
  M.~E.~Peskin and J.~D.~Wells,
  Phys.\ Rev.\ D {\bf 64}, 093003 (2001)
  [hep-ph/0101342].

\bibitem{Gerard:2007kn} 
  J.-M.~Gerard and M.~Herquet,
  Phys.\ Rev.\ Lett.\  {\bf 98}, 251802 (2007)
  [hep-ph/0703051 [HEP-PH]].

\bibitem{Grimus:2008nb} 
  W.~Grimus, L.~Lavoura, O.~M.~Ogreid and P.~Osland,
  Nucl.\ Phys.\ B {\bf 801}, 81 (2008)
  [arXiv:0802.4353 [hep-ph]].

\bibitem{Kanemura:2011sj} 
  S.~Kanemura, Y.~Okada, H.~Taniguchi and K.~Tsumura,
  Phys.\ Lett.\ B {\bf 704}, 303 (2011)
  [arXiv:1108.3297 [hep-ph]].







\bibitem{Barbieri:2006dq} 
  R.~Barbieri, L.~J.~Hall and V.~S.~Rychkov,
  Phys.\ Rev.\ D {\bf 74}, 015007 (2006)
  [hep-ph/0603188].
  
\bibitem{Baak:2014ora} 
  M.~Baak {\it et al.}  [Gfitter Group Collaboration],
  Eur.\ Phys.\ J.\ C {\bf 74}, 3046 (2014)
  [arXiv:1407.3792 [hep-ph]].

\bibitem{sirlin} 
  A.~Sirlin,
  Phys.\ Rev.\ D {\bf 22}, 971 (1980).
  

\bibitem{Hollik_EW}
  W.~F.~L.~Hollik,
  Fortsch.\ Phys.\  {\bf 38}, 165 (1990).

\bibitem{Logan:2000iv} 
  H.~E.~Logan and U.~Nierste,
  Nucl.\ Phys.\ B {\bf 586}, 39 (2000)
  [hep-ph/0004139].

\bibitem{Li:2014fea} 
  X.~Q.~Li, J.~Lu and A.~Pich,
  JHEP {\bf 1406}, 022 (2014)
  [arXiv:1404.5865 [hep-ph]].



\bibitem{Pich:2013lsa} 
  A.~Pich,
  Prog.\ Part.\ Nucl.\ Phys.\  {\bf 75}, 41 (2014)
  [arXiv:1310.7922 [hep-ph]].

\bibitem{PDG2014} 
K.A.~Olive et al.~(Particle Data Group), Chin.~Phys.~C, 38, 090001 (2014). 

\bibitem{Amhis:2014hma} 
  Y.~Amhis {\it et al.}  [Heavy Flavor Averaging Group (HFAG) Collaboration],
  arXiv:1412.7515 [hep-ex].

\bibitem{Staub:2013tta} 
  F.~Staub,
  Comput.\ Phys.\ Commun.\  {\bf 185}, 1773 (2014)
  [arXiv:1309.7223 [hep-ph]].


\bibitem{Gorczyca:2011rs} 
  B.~Gorczyca and M.~Krawczyk,
  Acta Phys.\ Polon.\ B {\bf 42}, 2229 (2011)
  [Erratum-ibid.\ B {\bf 43}, 481 (2012)]
  [arXiv:1112.4356 [hep-ph]];

  B.~Gorczyca and M.~Krawczyk,
  arXiv:1112.5086 [hep-ph].



\bibitem{Eriksson:2009ws} 
  D.~Eriksson, J.~Rathsman and O.~Stal,
  Comput.\ Phys.\ Commun.\  {\bf 181}, 189 (2010)
  [arXiv:0902.0851 [hep-ph]];

  J.~Rathsman and O.~Stal,
  PoS CHARGED {\bf 2010}, 034 (2010)
  [arXiv:1104.5563 [hep-ph]].


\bibitem{Koide} 
 H.~Fusaoka and Y.~Koide,
 Phys.\ Rev.\ D {\bf 57}, 3986 (1998). 


\bibitem{calchep}
  A.~Pukhov, E.~Boos, M.~Dubinin, V.~Edneral, V.~Ilyin, D.~Kovalenko, A.~Kryukov and V.~Savrin {\it et al.},
  hep-ph/9908288.


\bibitem{cteq} 
  J.~Pumplin, D.~R.~Stump, J.~Huston, H.~L.~Lai, P.~M.~Nadolsky and W.~K.~Tung,
  JHEP {\bf 0207}, 012 (2002).






\bibitem{Kanemura:2011kx} 
  S.~Kanemura, K.~Tsumura and H.~Yokoya,
  Phys.\ Rev.\ D {\bf 85}, 095001 (2012)
  [arXiv:1111.6089 [hep-ph]];

  S.~Kanemura, K.~Tsumura and H.~Yokoya,
  arXiv:1201.6489 [hep-ph].



 \bibitem{snowmass} 
   S.~Dawson, A.~Gritsan, H.~Logan, J.~Qian, C.~Tully, R.~Van Kooten, A.~Ajaib and A.~Anastassov {\it et al.},
   arXiv:1310.8361 [hep-ex].


\bibitem{fingerprint} 
  S.~Kanemura, K.~Tsumura, K.~Yagyu and H.~Yokoya,
  Phys.\ Rev.\ D {\bf 90}, 075001 (2014) 
  [arXiv:1406.3294 [hep-ph]].

\bibitem{yukawa_loop} 

  S.~Kanemura, M.~Kikuchi and K.~Yagyu,
  Phys.\ Lett.\ B {\bf 731}, 27 (2014)
  [arXiv:1401.0515 [hep-ph]];

  S.~Kanemura, M.~Kikuchi and K.~Yagyu,
  arXiv:1502.07716 [hep-ph].







\bibitem{Aad:2015vsa} 
  G.~Aad {\it et al.}  [ATLAS Collaboration],
  arXiv:1501.04943 [hep-ex].

\bibitem{CMS:2014ega} 
  CMS Collaboration [CMS Collaboration],
  CMS-PAS-HIG-14-009.


\bibitem{Aad:2014eha} 
  G.~Aad {\it et al.}  [ATLAS Collaboration],
  Phys.\ Rev.\ D {\bf 90}, no. 11, 112015 (2014)
  [arXiv:1408.7084 [hep-ex]].

\bibitem{Peskin:2012we} 
  M.~E.~Peskin,
  arXiv:1207.2516 [hep-ph].




  

\bibitem{Ohl:1995kr} 
 T.~Ohl,
 Comput.\ Phys.\ Commun.\  {\bf 90}, 340 (1995)
 [hep-ph/9505351].

\bibitem{Djouadi:2005gj} 
  A.~Djouadi,
  Phys.\ Rept.\  {\bf 459}, 1 (2008)
  [hep-ph/0503173] and references there in.


\end{thebibliography}
\end{document}